\documentclass[longauth]{aa}  
\usepackage{threeparttable}
\usepackage{graphicx}
\usepackage{txfonts}
\usepackage{xspace}
\usepackage{natbib,twoopt}
\usepackage[breaklinks=true]{hyperref} 
\bibpunct{(}{)}{;}{a}{}{,} 
\makeatletter
\newcommandtwoopt{\citeads}[3][][]{\href{http://adsabs.harvard.edu/abs/#3}%
{\def\hyper@linkstart##1##2{}%
\let\hyper@linkend\@empty\citealp[#1][#2]{#3}}}
\newcommandtwoopt{\citepads}[3][][]{\href{http://adsabs.harvard.edu/abs/#3}%
{\def\hyper@linkstart##1##2{}
\let\hyper@linkend\@empty\citep[#1][#2]{#3}}}
\newcommandtwoopt{\citetads}[3][][]{\href{http://adsabs.harvard.edu/abs/#3}%
{\def\hyper@linkstart##1##2{}
\let\hyper@linkend\@empty\citet[#1][#2]{#3}}}
\newcommandtwoopt{\citeyearads}[3][][]%
{\href{http://adsabs.harvard.edu/abs/#3}
{\def\hyper@linkstart##1##2{}%
\let\hyper@linkend\@empty\citeyear[#1][#2]{#3}}}
\makeatother
\newcommand{\juliet}{{\sc \tt juliet}\xspace}
\newcommand{\batman}{{\sc \tt batman}\xspace}
\newcommand{\celerite}{{\sc \tt celerite}\xspace}
\newcommand{\george}{{\sc \tt george}\xspace}
\newcommand{\radvel}{{\sc \tt radvel}\xspace}
\newcommand{\dynesty}{{\sc \tt dynesty}\xspace}
\newcommand{\e}[1]{{\times10^{#1}}}
\newcommand{\orcid}[1]{\protect\href{https://orcid.org/#1}{\protect\includegraphics[width=8pt]{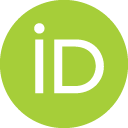}}}

\begin{document} 

\title{TOI-269~b: An eccentric sub-Neptune transiting a M2 dwarf revisited with ExTrA}
   
   \author{M.~Cointepas\inst{\ref{grenoble},\ref{geneva}}
   \thanks{Corresponding author:\texttt{marion.cointepas@univ-grenoble-alpes.fr}}
        \and J.M.~Almenara\orcid{0000-0003-3208-9815}\inst{\ref{grenoble}}
        \and X.~Bonfils\orcid{0000-0001-9003-8894}\inst{\ref{grenoble}}
        \and F.~Bouchy\inst{\ref{geneva}}
        \and N.~Astudillo-Defru\orcid{0000-0002-8462-515X}\inst{\ref{concepcion}}
        \and F.~Murgas\inst{\ref{grenoble},\ref{iac},\ref{ull}}
        \and J.F.~Otegi\inst{\ref{zurich},\ref{geneva}}
        \and A.~Wyttenbach\orcid{0000-0001-9003-7699}\inst{\ref{grenoble}}
        \and D.R.~Anderson\inst{\ref{keele},\ref{warwick1},\ref{warwick2}}
        \and \'E.~Artigau\inst{\ref{montreal}}
        \and B.L.~Canto~Martins\inst{\ref{brazil}}
        \and D.~Charbonneau\inst{\ref{harvard}}
        \and K.A.~Collins\inst{\ref{harvard}}
        \and K.I.~Collins\inst{\ref{mason}}
        \and J-J.~Correia\inst{\ref{grenoble}}
        \and S.~Curaba\inst{\ref{grenoble}}
        \and A.~Delboulb\'e\inst{\ref{grenoble}}
        \and X.~Delfosse\inst{\ref{grenoble}}
        \and R.F.~D\'iaz\inst{\ref{argentina}}
        \and C.~Dorn\inst{\ref{zurich}}
        \and R.~Doyon\orcid{0000-0001-5485-4675}\inst{\ref{montreal2}}
        \and P.~Feautrier\inst{\ref{grenoble}}
        \and P.~Figueira\inst{\ref{ESO},\ref{porto1}}
        \and T.~Forveille\inst{\ref{grenoble}}
        \and G.~Gaisne\inst{\ref{grenoble}}
        \and T.~Gan\inst{\ref{tsinghua}}
        \and L.~Gluck\inst{\ref{grenoble}}
        \and R.~Helled\orcid{0000-0001-5555-2652}\inst{\ref{zurich}}
        \and C.~Hellier\inst{\ref{keele}}
        \and L.~Jocou\inst{\ref{grenoble}}
        \and P.~Kern\inst{\ref{grenoble}}
        \and S.~Lafrasse\inst{\ref{grenoble}}
        \and N.~Law\inst{\ref{chapelhill}}
        \and I.C.~Le\~ao\inst{\ref{brazil}}
        \and C.~Lovis\inst{\ref{geneva}}
        \and Y.~Magnard\inst{\ref{grenoble}}
        \and A.W.~Mann\inst{\ref{chapelhill}}
        \and D.~Maurel\inst{\ref{grenoble}}
        \and J.R.~de~Medeiros\inst{\ref{brazil}}
        \and C.~Melo\inst{\ref{ESO}}
        \and T.~Moulin\inst{\ref{grenoble}}
        \and F.~Pepe\inst{\ref{geneva}}
        \and P.~Rabou\inst{\ref{grenoble}}
        \and S.~Rochat\inst{\ref{grenoble}}
        \and D.R.~Rodriguez\orcid{0000-0003-1286-5231}\inst{\ref{stsi}} 
        \and A.~Roux\inst{\ref{grenoble}}
        \and N.C.~Santos\inst{\ref{porto1},\ref{porto2}}
        \and D.~S\'egransan\inst{\ref{geneva}}
        \and E.~Stadler\inst{\ref{grenoble}}
        \and E.B.~Ting\orcid{0000-0002-8219-9505}\inst{\ref{ames}}
        \and J.D.~Twicken\orcid{0000-0002-6778-7552}\inst{\ref{seti},\ref{ames}}
        \and S.~Udry\inst{\ref{geneva}}
        \and W.C.~Waalkes\orcid{0000-0002-8961-0352}\inst{\ref{boulder}}
        \and R.G.~West\inst{\ref{warwick1},\ref{warwick2}}
        \and A.~W\"{u}nsche\inst{\ref{grenoble}}
        \and C.~Ziegler\inst{\ref{dunlap}}
        \and G.~Ricker\inst{\ref{mit}}
        \and R.~Vanderspek\inst{\ref{mit}}
        \and D.W.~Latham\inst{\ref{harvard}}
        \and S.~Seager\inst{\ref{mit},\ref{mit2},\ref{mit3}}
        \and J.~Winn\inst{\ref{princeton}}
        \and J.M.~Jenkins\orcid{0000-0002-4715-9460}\inst{\ref{ames}}
}

      \institute{
        Univ. Grenoble Alpes, CNRS, IPAG, F-38000 Grenoble, France\label{grenoble}
        \and Observatoire de Gen\`eve, Département d’Astronomie, Universit\'e de Gen\`eve, Chemin Pegasi 51b, 1290 Versoix, Switzerland\label{geneva}
        \and Departamento de Matem\'{a}tica y F\'{i}sica Aplicadas, Universidad Cat\'{o}lica de la Sant\'{i}sima Concepci\'{o}n, Alonso de Rivera 2850, Concepci\'{o}n, Chile\label{concepcion}
        \and Astrophysics Group, Keele University, Staffordshire ST5 5BG, UK\label{keele}
        \and Department of Physics and Astronomy, The University of North Carolina at Chapel Hill, Chapel Hill, NC 27599-3255, USA\label{chapelhill}
        \and Dunlap Institute for Astronomy and Astrophysics, University of Toronto, 50 St. George Street, Toronto, Ontario M5S 3H4, Canada\label{dunlap}
        \and Department of Physics, University of Warwick, Coventry, UK\label{warwick1}
        \and Centre for Exoplanets and Habitability, University of Warwick, Gibbet Hill Road, Coventry, UK\label{warwick2}
        \and Center for Astrophysics \textbar \ Harvard \& Smithsonian, 60 Garden Street, Cambridge, MA 02138, USA\label{harvard}
        \and George Mason University, 4400 University Drive, Fairfax, VA, 22030 USA\label{mason}
        \and Department of Astronomy and Tsinghua Centre for Astrophysics, Tsinghua University, Beijing 100084, China\label{tsinghua}
        \and Department of Astrophysical and Planetary Sciences, University of Colorado, Boulder, CO 80309, USA\label{boulder}
        \and Institute for Computational Science, University of Zurich, Winterthurerstr. 190, CH-8057 Zurich, Switzerland \label{zurich}
        \and D\'epartement de physique, Universit\'e de Montr\'eal, 2900 boul.
        \'Edouard-Montpetit, Montr\'eal, QC, H3C 3J7, Canada\label{montreal}
        \and Universit\'e de Montr\'eal, D\'epartement de Physique \& Institut de Recherche sur les Exoplan\`etes, Montr\'eal, QC H3C 3J7, Canada\label{montreal2}
        \and International Center for Advanced Studies (ICAS) and ICIFI(CONICET), ECyT-UNSAM, Campus Miguelete, 25 de Mayo y Francia(1650), Buenos Aires, Argentina\label{argentina}
        \and European Southern Observatory, Alonso de C\'ordova 3107, Vitacura,
        Regi\'on Metropolitana, Chile\label{ESO}
        \and Instituto de Astrof\'isica e Ci\^encias do Espa\c{c}o,
        Universidade do Porto, CAUP, Rua das Estrelas, 4150-762 Porto, Portugal\label{porto1}
        \and Departamento de F\'isica e Astronomia, Faculdade de Ci\^encias,
        Universidade do Porto, Rua do Campo Alegre, 4169-007 Porto, Portugal\label{porto2}
        \and Departamento de F\'isica, Universidade Federal do Rio Grande do
        Norte, 59072-970 Natal, RN, Brazil\label{brazil}
        \and Instituto de Astrofísica de Canarias (IAC), E-38200 La Laguna, Tenerife, Spain\label{iac}
        \and Dept. Astrofísica, Universidad de La Laguna (ULL), E-38206 La Laguna, Tenerife, Spain\label{ull}
        \and Department of Physics and Kavli Institute for Astrophysics and Space Research, Massachusetts Institute of Technology, Cambridge, MA 02139, USA\label{mit}
        \and Department of Earth, Atmospheric and Planetary Sciences, Massachusetts Institute of Technology, Cambridge, MA 02139, USA\label{mit2}
        \and Department of Aeronautics and Astronautics, MIT, 77 Massachusetts Avenue, Cambridge, MA 02139, USA\label{mit3}
        \and Department of Astrophysical Sciences, Princeton University, NJ 08544, USA\label{princeton}
        \and NASA Ames Research Center, Moffett Field, CA 94035, USA\label{ames}
        \and SETI Institute, Mountain View, CA 94043, USA\label{seti}
        \and Space Telescope Science Institute, 3700 San Martin Drive, Baltimore, MD 21218, USA\label{stsi}
}

\date{Received ; Accepted}

\abstract
{We present the confirmation of a new sub-Neptune close to the transition between super-Earths and sub-Neptunes transiting the M2 dwarf TOI-269 (TIC 220479565, $V = 14.4$ mag, $J = 10.9$ mag, $R_{\star}=0.40~R_{\odot}$, $M_{\star}=0.39~M_{\odot}$, d = 57 pc). The exoplanet candidate has been identified in multiple TESS sectors, and  validated with high-precision spectroscopy from HARPS and ground-based photometric follow-up from ExTrA and LCO-CTIO. 
We determined mass, radius, and bulk density of the exoplanet by jointly modeling both photometry and radial velocities with \juliet. The transiting exoplanet has an orbital period of $P=3.6977104\pm0.0000037$ days, a radius of $2.77\pm0.12$ $R_{\oplus}$, and a mass of $8.8\pm1.4$ $M_{\oplus}$. Since TOI-269~b lies among the best targets of its category for atmospheric characterization, it would be interesting to probe the atmosphere of this exoplanet with transmission spectroscopy in order to compare it to other sub-Neptunes. With an eccentricity e = $0.425^{+0.082}_{-0.086}$, TOI-269~b has one of the highest eccentricities of the exoplanets with periods less than 10 days. The star being likely a few Gyr old, this system does not appear to be dynamically young. We surmise TOI-269~b may have acquired its high eccentricity as it migrated inward through planet-planet interactions. 
}


\maketitle
 
\section{Introduction}

Low-mass stars offer particular  advantages, and are particularly interesting, when looking for small and cool exoplanets. The transit depth and the radial-velocity semi-amplitude provoked by rocky exoplanets around M dwarfs is much higher than those caused by similar planets orbiting around larger stars of earlier spectral type, which makes such planets easier to detect. Even more importantly, such systems are also ideal targets for atmospheric characterization by transmission or thermal emission spectroscopy (e.g., \citealt{kempton2018}, \citealt{batalha2018}). Planets orbiting around M dwarfs are therefore important objects to obtain precise mass, radius, and bulk density measurements of transiting exoplanets smaller than Neptune, which are crucial to better understanding the so-called radius valley between the super-Earth and sub-Neptune populations (e.g., \citealt{fulton2017}, \citealt{mayo2018}, \citealt{cloutier2020}, and \citealt{hardegree2020}).\\

The Transiting Exoplanet Survey Satellite \citep[TESS,][]{ricker2015} is a NASA all-sky survey designed to detect transiting exoplanets orbiting nearby stars, with a specific focus on exoplanets smaller than Neptune. With its array of four cameras, the satellite has been delivering, since July 2018, both 2 min cadence photometry on pre-selected targets and full-frame images (FFIs) with a cadence of 30~min (no longer the case, however,  for the extended mission that began on  July 5, 2020, when TESS started collecting FFIs at 10 min cadence). Each pointing corresponds to a sector of the sky that is observed for a period of about 27 days. TESS is focusing on relatively bright stars, which is crucial for the confirmation of transiting candidates and to enable follow-up observations and characterization of confirmed exoplanets. 
TESS has identified more than 200 exoplanet candidates orbiting M dwarfs to date, including a few super-Earths with masses measured thanks to high-precision radial velocities (e.g., GJ357~b - \citealt{luque2019}; LTT3780~b - \citealt{cloutier2020b}).\\ 

The Exoplanets in Transits and their Atmospheres (ExTrA) facility \citep{bonfils2015} was developed to specifically perform from the ground high-precision spectro-photometry in the near-infrared on mid- to late M dwarf stars. ExTrA consists of a set of three 60 cm telescopes equipped with multi-object fiber positioners and a low-resolution near-IR spectrograph. ExTrA is being used to confirm TESS planet detections, refine transit parameters, and search for additional exoplanets in the same systems.\\

In this paper we report the discovery of a new transiting exoplanet around the M2 dwarf TOI-269, which was first detected as a candidate by TESS. The target was then confirmed with ground-based photometry and its mass measured thanks to HARPS radial velocities (RVs). Section 2 presents a detailed analysis of the stellar properties of TOI-269. Section 3 describes the observations and data used in this study, including a description of ExTrA. Section 4 presents the global analysis of the available data in order to constrain the planetary properties. Section~\ref{section.discussion} presents a discussion of our results, and Section 6 presents our conclusions. \\

\section{Stellar parameters}\label{section:stellarparameters}

TOI-269 is an M dwarf at a distance of $57.023\pm0.076$~pc (\citealt{gaia2018}, \citealt{lindegren2018}, and \citealt{bailer-jones2018}). The astrometry, photometry, and   stellar parameters are reported in Table~\ref{tab:stellar_params}.\\

\begin{table}[h]
      \caption[]{Stellar parameters for TOI-269 (TIC 220479565, UCAC4 180-005252, 2MASS J05032306-5410378, WISE J050323.09-541039.1, APASS 26149036).}
         \label{tab:stellar_params}
         \begin{tabular}{lcc}
            \hline
            \noalign{\smallskip}
            Parameter & \text{Value} & \text{Refs} \\
            \noalign{\smallskip}
            \hline
            \noalign{\smallskip}
            \textit{Astrometry} \\
            Right ascension (J2015.5), $\alpha$ & 05$^{\rm h}$03$^{\rm m}$23.11$^{\rm s}$ & 1,2 \\
            Declination (J2015.5), $\delta$ & $-$54$^{\rm o}$10'39.8'' & 1,2 \\
            Parallax, [mas] & $17.508\pm0.023$ & 1,2 \\
            Distance, d [pc] & $57.023\pm0.076$ & 1,2 \\
            Proper motion RA [mas/year] & $26.472\pm0.43$ & 1,2 \\
            Proper motion D [mas/year] & $120.750\pm0.43$ & 1,2 \\
            \noalign{\smallskip}
            \textit{Photometry} \\
            V [mag]& $14.37\pm0.11$ & 3 \\
            TESS magnitude [mag]& $12.2958\pm0.0074$ & 3 \\
            J [mag]& $10.909\pm0.026$ & 4 \\
            H [mag]& $10.304\pm0.022$ & 4 \\
            $K_s$ [mag]& $10.100\pm0.023$ & 4 \\
            \noalign{\smallskip}
            \textit{Stellar parameters} \\
            Spectral type & M2V & 5 \\
            $M_{K_s}$ [mag]& $6.320\pm0.023$ & 6 \\
            Effective temperature, $T_{\rm eff}$ [K] & $3514\pm70$ & 6,7 \\
            Surface gravity, log g [cgs] & $4.831\pm0.029$ & 6 \\
            Metallicity, [Fe/H] [dex] & $-0.29\pm0.12$ & 6,7 \\
            Stellar radius, $R_{\star}$ [$R_\odot$] & $0.398\pm0.012$ & 6 \\
            Stellar mass, $M_{\star}$ [$M_\odot$] & $0.3917\pm0.0095$ & 6 \\
            log $R'_{HK}$ & $-5.320\pm0.142$ & 6 \\
            \noalign{\smallskip}
            \hline
        \end{tabular}
        \begin{tablenotes}
        \small
        \item References : 1) \cite{gaia2018}, 2) \cite{lindegren2018}, 3) \cite{stassun2019}, 4) \cite{cutri2003}, 5) \cite{pecaut2013}, 6) This work, 7) \cite{yee2017}, 
        \end{tablenotes}
\end{table}

We derived the mass of the star with the empirical relationship between $M_{K_s}$ and $M_\star$ of \cite{mann2019} and using their {\sc \tt M\_-M\_K-} code\footnote{https://github.com/awmann/M\_-M\_K-}. An apparent magnitude of $K_s=10.100\pm0.023$ leads to an absolute magnitude of $M_{K_s}=6.320\pm0.023$ and a mass of $M_{\star}=0.3917\pm0.0095~M_\odot$. We used another empirical relation from \cite{mann2015} to estimate the radius of the star also from the value of the apparent magnitude and found $R_{\star}=0.398\pm0.012~R_\odot$, which is the same value as provided in the TESS Input Catalog \citep[TIC-v8;][]{stassun2019}.
With both mass and radius we infer a stellar density of $\rho_{\star}=8.76\pm0.87~g$/$cm^3$.
The analysis by Mann includes stars with close-to-solar metallicities. However, using {\sc \tt SpecMatch-Emp}\footnote{https://github.com/samuelyeewl/specmatch-emp} \citep{yee2017} and HARPS high-resolution spectra of TOI-269, we estimated that the star has a metallicity of ${\rm [Fe/H]} = -0.29\pm0.12$ \citep{yee2017}. To verify the impact of such sub-solar metallicity on the stellar parameters, we performed an analysis of the spectral energy distribution combined with stellar evolution model (Appendix~\ref{section:sed}). It leads to similar results for the stellar mass and radius. To be more conservative,  in the rest of this paper we use the relations in \citet{mann2015,mann2019} without taking into account the stellar metallicity. 
With {\sc \tt SpecMatch-Emp}, we also obtain an effective temperature $T_{\rm eff}=3514\pm70~K$ and radius $R_\star=0.41\pm0.04~R_\odot$. We report the $T_{\rm eff}$ value in Table~\ref{tab:stellar_params}, but discard the $R_\star$ value, which we found to be less precise than that of the \citet{mann2015} relationship.
An analysis of the WISE \citep{wright2010} data for TOI-269 shows no IR excess, pointing to the absence of circumstellar material around the star, at least in the WISE bands.

We estimated the age of the star using gyrochronology with the different rotation periods that we calculated  (see Sections 3.2.3 and 3.3). We used the relation from \cite{engle2011}, which determines the stellar age  through measures of the rotation period of the star. If we used the estimation of the rotation period from the HARPS spectra ($P_{\rm rot} = 69\pm15$~days), we obtained a stellar age of $4.19\pm0.97$~Gyr. We also calculated the age with the rotation period measured from WASP-South ($P_{\rm rot} = 35.7\pm2.0$~days) and found $2.16\pm0.27$~Gyr. At this point, we cannot conclude  which rotation period is the correct one, but it appears that the system is not newly formed. We confirmed these estimations using the relations from \citet{barnes2010}, \citet{kim2010}, and \citet{meibom2015}, which describe the rotational evolution of main sequence stars taking into account the stellar mass dependence, as we obtained an age of $5.8^{+2.4}_{-1.9}$~Gyr from the HARPS rotation period and $2.36^{+0.45}_{-0.35}$~Gyr from the WASP-South data.
In terms of activity, we found the star to be rather quiet and slowly rotating (see next section).

\section{Observations}

\subsection{TESS photometry}

\begin{figure}[h]
   \centering
   \includegraphics[width=0.5\textwidth]{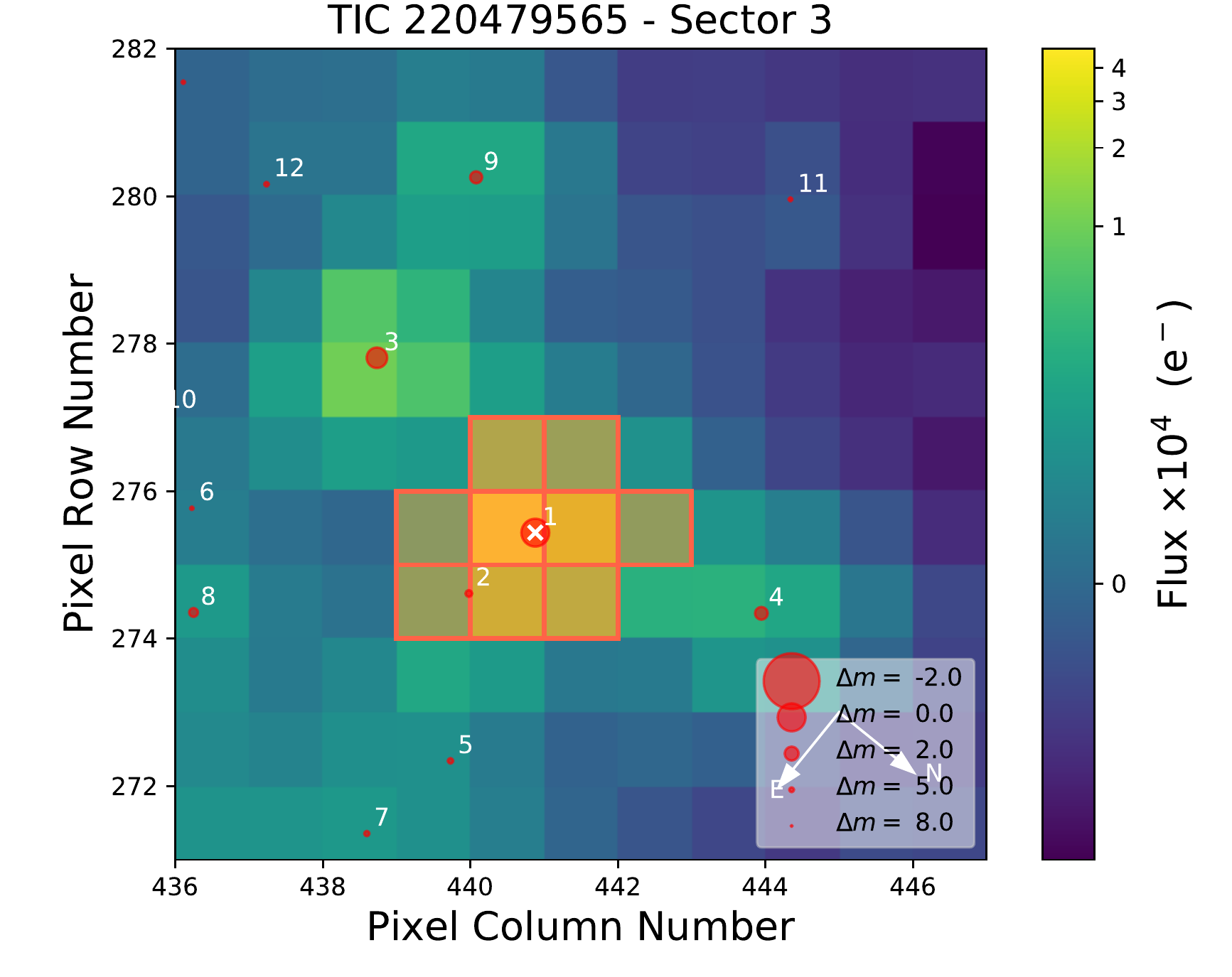}
      \caption{TESS target pixel file image of TOI-269 in Sector 3 \citep[created with {\sc \tt tpfplotter},][]{aller2020}. The electron counts are color-coded. The red bordered pixels are used in the simple aperture photometry. The size of the red circles indicates the Gaia DR2 magnitudes of all nearby stars.}
      \label{fig:tpf}
\end{figure}

TOI-269 was observed in six TESS sectors (3, 4, 5, 6, 10, and 13) from September 2018 to July 2019 with the two-minute cadence.
The TESS photometric data were processed by the NASA Ames Science Processing Operations Center \citep[SPOC;][]{jenkins2016}. The resulting Presearch Data Conditioning Simple Aperture Photometry (PDCSAP; \citealt{smith2012}, \citealt{stumpe2012,stumpe2014}) light curve of TOI-269 was corrected for dilution by known contaminating sources in the TESS aperture.
Given the large TESS pixel size of 21", it was essential to verify that no visually close-by targets were present that could affect the depth of the transit and to check for a contaminating eclipsing binary. A plot of the target pixel file (TPF) and the aperture mask that is used for the simple aperture photometry (SAP), generated with {\sc \tt tpfplotter} \citep{aller2020}, is shown in Fig.~\ref{fig:tpf}. We can see that one star overlaps the TESS aperture, but its faintness results in minimal dilution of the TESS light curve. The CROWDSAP value in the TESS data (which is the ratio of target flux to total flux in the aperture) ranges from 0.9829 to 0.9932 in all TESS sectors used for this analysis. This value corresponds to the $\sim$~1\% dilution expected from the star visible in the TESS aperture with a magnitude difference of 5 with TOI-269. \\

A potential transit signal with 3.698-day period was identified in the SPOC transit search (\citealt{jenkins2002}, \citealt{jenkins2010,jenkins2020}) of the TOI-269 light curve. This was promoted to TOI planet candidate status and designated as TOI-269.01 by the TESS Science Office based on the SPOC data validation results; the results  indicate that the 4900 ppm signature based on 35 individual transits was consistent with that of a transiting planet (\citealt{twicken2018}, \citealt{li2019}). 
The difference image centroids analysis performed in the data validation (DV) report shows that the source of the transits is within 1 arcsec of the target, and furthermore that the DV reports are very clean;  the transiting planet’s signature passes all of DV tests, including the odd/even transit depth test, the ghost diagnostic, and the statistical bootstrap test. These tests are described in \cite{twicken2018} and can exclude a false positive created by the second star in TESS aperture.\\

A recent work by \cite{canto2020}, which reports a search for rotation period among 1000 TESS TOIs using fast Fourier transform, Lomb–Scargle, and wavelet techniques, accompanied by a rigorous visual inspection shows that TOI-269 does not indicate any sign of chromospherical activity all along the observed time ranges, and that it is among the quietest stars in the sample.

\subsection{Ground-based photometry}

\subsubsection{ExTrA photometry}
ExTrA is a new facility composed of three 60 cm telescopes that are located at La Silla Observatory \citep{bonfils2015}. It is dedicated to searching for exoplanets transiting nearby M dwarfs with near-infrared photometry.
The instrument relies on a new approach that involves combining photometry with spectroscopic information in order to mitigate the disruptive effect of Earth’s atmosphere, as well as effects introduced by instruments and detectors. On each telescope five field units (FUs) are used to collect the light from the main target and four selected comparison stars. The FUs are composed of fibers accurately positioned that all feed a single near-infrared spectrograph with low spectral resolution ($R\sim200$ or $R\sim20$) and that covers the 0.9--1.6~$\mu$m range. The wide spectral domain enables the simultaneous collection of more photons which increases the sensitivity. ExTrA has the advantage of multiplexing, meaning that we can use a single spectrograph (so a single detector) to record the light from several telescopes. Although all three telescopes can observe the same field,  most of the time they will observe different fields in order to maximize the transit search efficiency. For now all the telescopes are observing the same target in order to better understand the performance and limitations of the instrument. \\
On each telescope each FU is  composed of two "buttons": the science button and the centering button. The centering button is composed of a bundle of 19 fibers in a hexagonal arrangement, used to calculate the position of the stars accurately, and then apply a relative offset to center the star in the science button. Each science button is composed of two channels, one with an aperture of 8" (optimized for bright stars), and a second one  with an aperture of 4"  in order to reduce the background contribution (optimized for faint stars). Finally, each channel has two fibers: one to collect the light of the star and another to collect the sky background close to the star. Thus, the spectrograph  images the fiber bundles in centering mode, and with a different light path disperses the science fibers using a prism. \\

We observed TOI-269 with an 8" fiber aperture using the low-resolution mode ($R\sim20$). We used Gaia DR2 4770833527616320512 as a guiding star to correctly point our target and comparison stars. The five stars we observed for that specific field of view were 2MASS J05021375-5416532 on the field unit 1 (FU1), 2MASS J05032583-5353188 on the FU3, 2MASS J05020219-5351064 on the FU4, 2MASS J05001046-5407021 on the FU5, and finally our target star, which was positioned on the FU2. The four comparison stars come from the 2MASS catalog \citep{2mass}. The field of view of TOI-269 used for each night of observations with ExTrA is presented in the Appendix in Fig.~\ref{fig:field}. At low resolution the spectrum of the star only covers 43~pixels of the 640 available in the detector in the spectral dimension. We first subtracted the sky flux and normalized by the flat. We made sure by looking at the absolute flux of all the stars that the night was clear and that we did not have any strong flux loss that could cause issues in the photometry. We created a template of all the exposures to remove any outliers. We distributed a weight (proportional to the variance of flux versus time) for each pixel/wavelength in order to mitigate the effect of the Earth's atmosphere, especially in the water bands. We then obtained the differential photometry for each star (our target and the four comparison stars) by normalizing with a template that includes the photometry of the different stars. The photometry is still affected by some systematic effects that are still under investigation. One of them may come from the fact that there are some differential motions of the stars during the observing sequence which introduce differential centering on the fibers aperture. A decorrelation with the position of the star inside the fiber is not possible because we cannot monitor the XY position of the star inside the fiber aperture during an acquisition. This problem is being investigated by observing the same field with the centering bundle only during a whole night, and then applying the measured offsets on all subsequent nights.
For this work, we only present the light curve of our main target TOI-269 and we removed systematic trends applying Gaussian processes that allow us to model non-white noise effects for which we do not have a good model at hand. TOI-269~b transits were observed with ExTrA on December 18, 2019, December 29, 2019, and March 1, 2020, with an exposure time of 60~seconds. Telescope 1 was not in operation during the first two nights. In total, seven light curves of TOI-269 were obtained with ExTrA.\\

\subsubsection{LCOGT Photometry}
We observed full transits of TOI-269~b in Sloan $i'$-band on UTC 2019 March 02 (and respectively UTC 2020 October 13, only used in Section~\ref{section.ttvs} as the global analysis had already been done at this time) from the LCOGT \citep{Brown:2013} 1.0\,m network nodes at Cerro Tololo Inter-American Observatory (CTIO) (and Siding Spring Observatory). We also observed an ingress with nearly full transit coverage in ${\rm I_c}$-band on UTC 2019 April 07 from the LCOGT 1.0\,m network node at CTIO. We used the {\tt TESS Transit Finder}, which is a customized version of the {\tt tapir} software package \citep{Jensen:2013} to schedule our transit observations. The $4096\times4096$ LCOGT SINISTRO cameras have an image scale of 0.389$\arcsec$ per pixel, resulting in a $26\arcmin\times26\arcmin$ field of view. The images were calibrated by the standard LCOGT {\tt BANZAI} pipeline \citep{McCully:2018}, and photometric data were extracted with {\tt AstroImageJ} \citep{Collins:2017}. The images were mildly defocused and have typical stellar point spread functions with a full width half maximum (FWHM) of $\sim2\farcs 3$;  circular apertures with radius $\sim4\farcs3$ were used to extract the differential photometry. 

\subsubsection{WASP-South photometry}
\begin{figure*}
   \centering
   \includegraphics[width=0.98\textwidth]{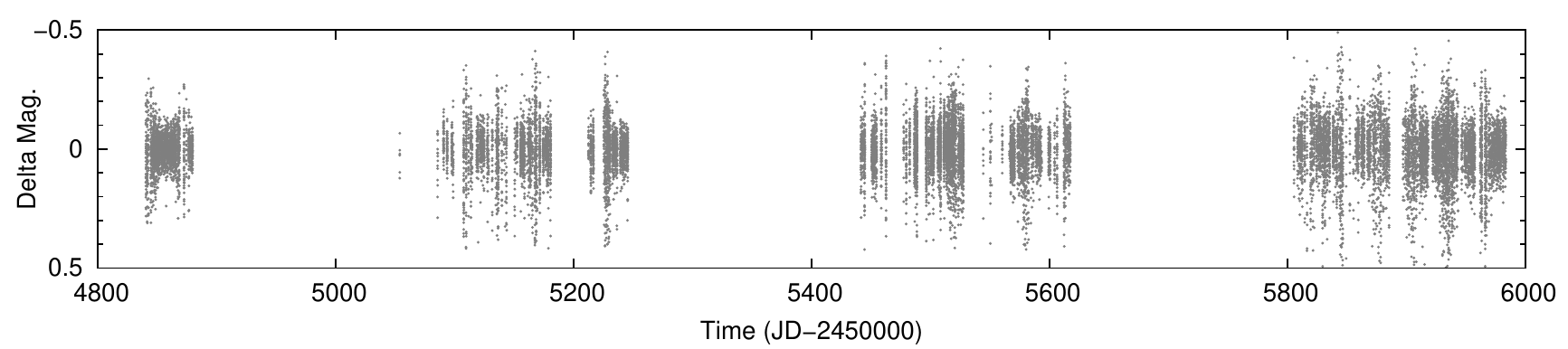}
      \caption{WASP-South light curve of TOI-269 showing the time sampling.}
      \label{fig:wasp_lc}
\end{figure*}

\begin{figure}[h]
   \centering
   \includegraphics[width=0.5\textwidth]{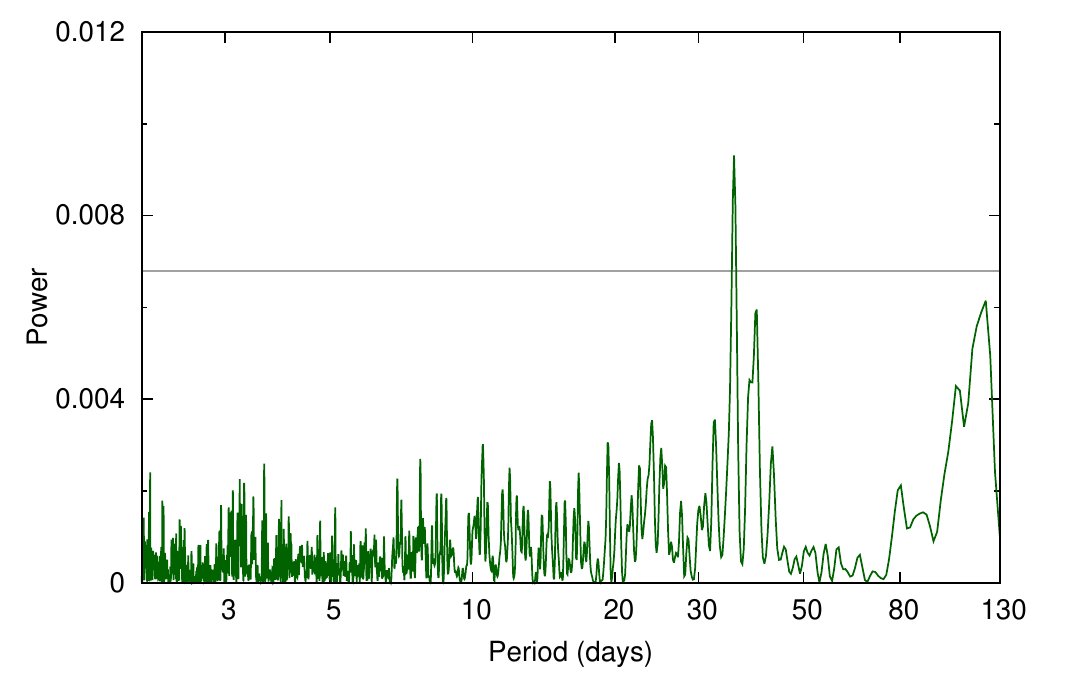}
      \caption{Periodogram of the WASP-South light curves of TOI-269 from 2009--2011, showing a possible rotational modulation with a period of 36 days. The horizontal line corresponds to the estimated 1\% false alarm level.}
      \label{fig:wasp}
\end{figure}

The field of TOI-269 was observed by WASP-South (an array of eight cameras using 200 mm, f/1.8 lenses backed by $2048\times2048$ CCDs; \citealt{2006PASP..118.1407P}) over the three consecutive years, 2009, 2010, and 2011. Observations spanned 180~days in each year, accumulating a total of 18\,100 photometric data points (Fig.~\ref{fig:wasp_lc}). At a magnitude of $V$ = 14.4, TOI-269 is at the faint end of the WASP range, but the field is relatively sparse, and the next-brightest star in the 48 arcsec extraction aperture is 4 magnitudes fainter. \\
Combining all the data, and searching for a rotational modulation using the methods presented in \citet{2011PASP..123..547M}, we find a possible periodicity  of 35.7 $\pm$ 2.0 d, where the error allows for possible phase changes over the observation span (Fig.~\ref{fig:wasp}). The modulation has an amplitude of 10~mmag, and an estimated false alarm probability of less than 1\%. \\

\subsection{Radial velocities:  HARPS}
We obtained 81 spectra of TOI-269 with the High Accuracy Radial velocity Planet Searcher \citep[HARPS;][]{mayor2003} echelle spectrograph at the ESO 3.6m telescope at La Silla Observatory in Chile. The HARPS optical spectrograph has a resolution of R=115,000 and is stabilized in pressure and temperature, in order to achieve a sub-m/s precision. The observations span 275 days and were taken between UT January 18, 2019, and October 20, 2019 (Prog-ID 1102.C-0339(A)).\\
The HARPS observations were acquired without simultaneous wavelength calibration and with an exposure time of 1,800 seconds, although seven of them have exposures of 1,200 or 2,400 seconds, resulting in a total open shutter time of 41.2 hours. These data were reduced with the Data Reduction Software\footnote{http://www.eso.org/sci/facilities/lasilla/instruments/harps/doc/DRS.pdf}. Radial velocities were computed through maximum likelihood between individual spectra and a master stellar spectrum following the recipes described in \citet{Astudillo-Defru2017b}. Reduced spectra have a median signal-to-noise ratio of $\sim$13 at 650~nm, resulting in a median radial velocity precision of 7.5~m/s and presenting a dispersion of 8.5~m/s.\\

We also performed an analysis with HARPS data using the DACE platform \footnote{https://dace.unige.ch} \citep{dace2019} that can be used to compute periodograms for parameters available in RVs data and to represent correlations between different indicators. We did not find any sign of a strong stellar activity in the different indicators (CCF-FWHM, S-index, Na-index) derived from the spectra except in H$\alpha$ with a peak around 54~days. The low value of $logR'_{HK} = -5.32\pm0.14$, derived from the HARPS combined spectra, indicates a chromospherically inactive star with likely a long rotation period estimated to $P_{\rm rot} = 69\pm15$~days, according to \cite{astudillo2017}.\\
This value is compatible with 2$\sigma$ to the WASP-South value, but we could also imagine that the signal found in the WASP data comes from two stellar spots at the surface of the star and could correspond to $P_{\rm rot}/2$. \\

\subsection{SOAR speckle imaging}
\begin{figure}[h]
   \centering
   \includegraphics[width=0.5\textwidth]{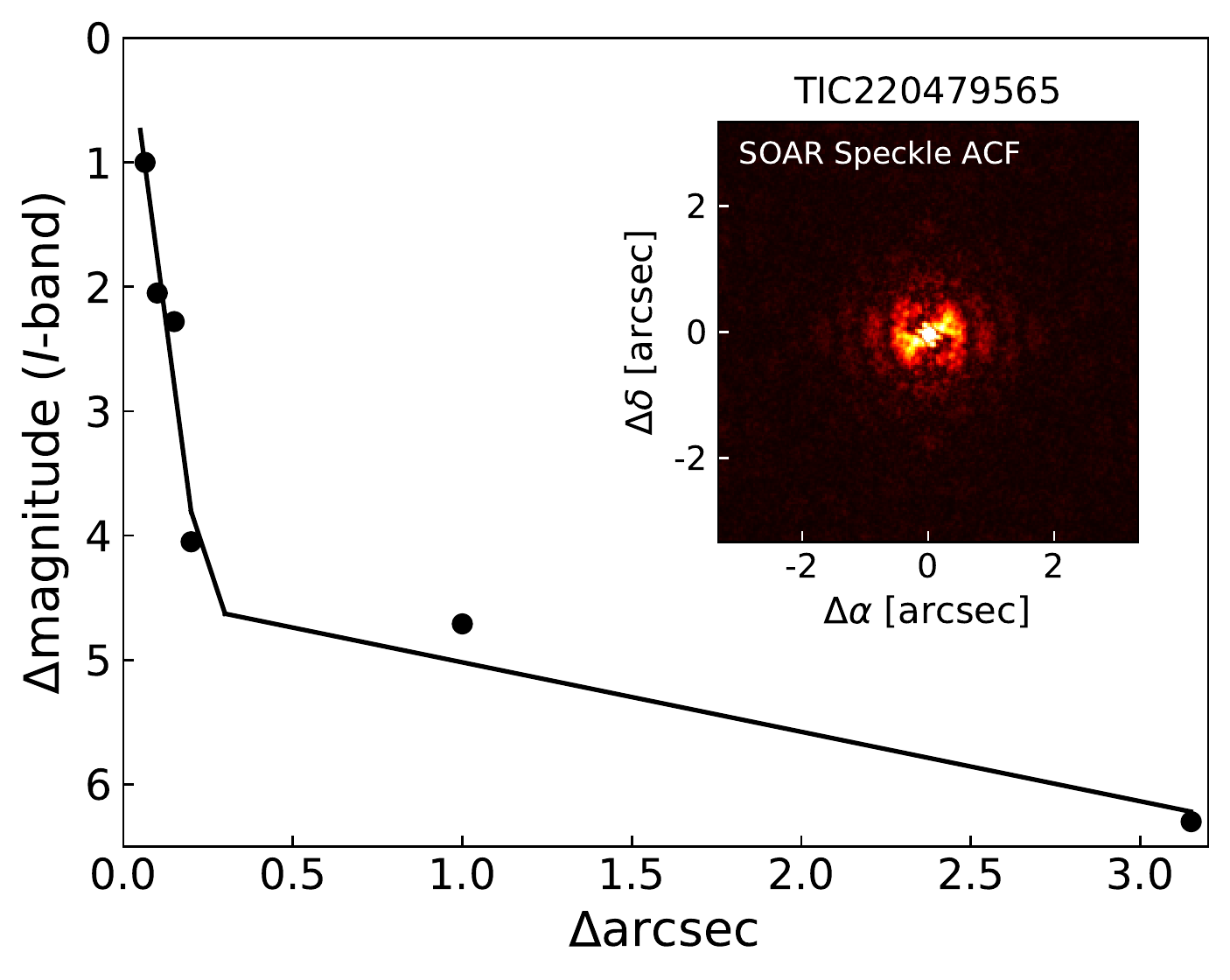}
      \caption{I-band 5$\sigma$ contrast curve from SOAR speckle imaging of TOI-269 (TIC 220479565). The inset depicts the corresponding speckle auto-correlation function.}
     \label{fig:speckle}
\end{figure}

High angular resolution imaging is needed to search for nearby sources that could contaminate the TESS photometry, resulting in an underestimated planetary radius, or could be the source of astrophysical false positives, such as background eclipsing binaries. We searched for stellar companions to TOI-269 with speckle imaging on the 4.1 m Southern Astrophysical Research (SOAR) telescope \citep{tokovinin2018} on 18 February 2019 UT, observing in Cousins $I$ band, a  visible band-pass similar to TESS. More details of the observations are available in \cite{ziegler2020}. The 5$\sigma$ detection sensitivity and speckle auto-correlation functions from the observations are shown in Fig.~\ref{fig:speckle}. No nearby stars were detected within 3\arcsec of TOI-269 in the SOAR observations.

\section{Analysis and modeling}
We used the software package \juliet \citep{espinoza2019} to model separately and jointly the photometric and radial-velocity data. The algorithm is built on many publicly available tools for the modeling of transits \citep[\batman,][]{kreidberg2015}, RVs \citep[\radvel,][]{fulton2018}, and Gaussian processes (GPs) (\george, \citealt{ambikasaran2015}; \celerite, \citealt{foreman2017}). In order to compare different models, \juliet efficiently computes the Bayesian evidence (lnZ) using \dynesty \citep{speagle2020}, a python package to estimate Bayesian posteriors and evidence using nested sampling methods. Nested sampling algorithms sample directly from the given priors instead of starting off with an initial parameter vector around a likelihood maximum found via optimization techniques, as is done in common sampling methods. During our different analysis, we made sure that we had enough live points given the number of free parameters so that we would not miss peaks in the parameter space. We started the analysis using only TESS photometry, and then HARPS RVs measurements in order to constrain the priors and use them for a joint analysis of all our data.\\

\subsection{TESS photometry}
\begin{figure*}[h]
    \centering
    \includegraphics[width=1.0\textwidth]{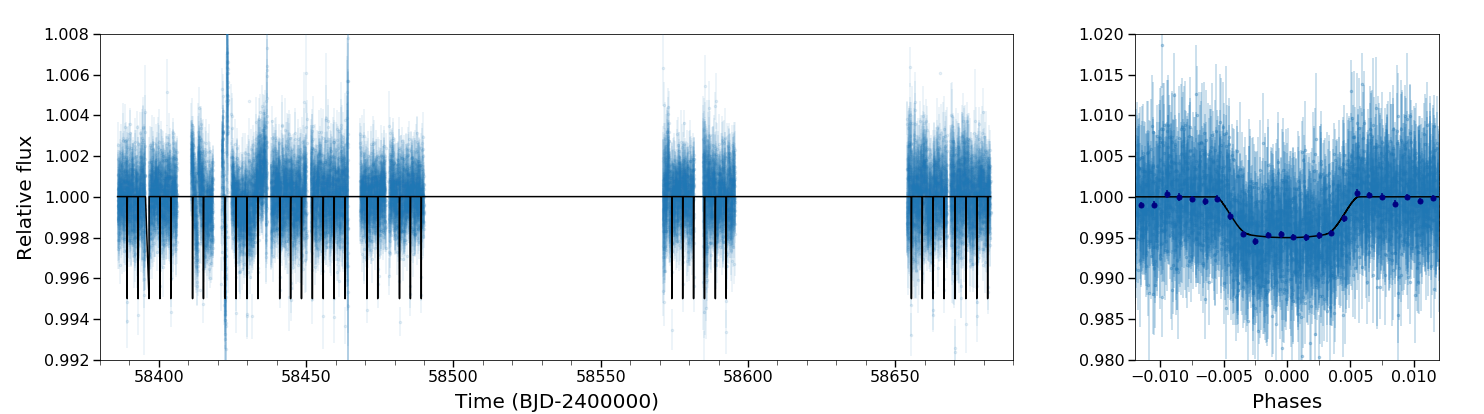}
    \caption{Modeling of the TESS data. Left panel: TESS photometry time series (20 min binned data) from Sectors 3, 4, 5, 6, 10, and 13 along with the median of 1000 randomly chosen posterior samples (solid black line). Right panel: Phase-folded light curve to the period of the planet (dark blue points correspond to $\sim$~5 min binned data).}
    \label{fig:tessall}
\end{figure*}

First,  using \juliet, we modeled the TESS PDCSAP light curve where our planet candidate was initially detected. The transit model fits the stellar density $\rho_{\star}$ along with the planetary and jitter parameters. We chose the priors of the orbital parameters from ExoFOP, except for the stellar density that we calculated in section 2. We adopted a few parametrization modifications when dealing with the transit photometry. We assigned a quadratic limb-darkening law for TESS, as shown to be appropriate for space-based missions \citep{espinoza2015}, which was then  parameterized with the uniform sampling scheme $(q_1,q_2)$ introduced by \cite{kipping2013}. Additionally, rather than fitting directly for the planet-to-star radius ratio ($p=R_p/R_{\star}$) and the impact parameter of the orbit ($b=a/R_{\star}\cos{i}$), \juliet used the parameterization introduced in \cite{espinoza2018} and fit for the parameters $r_1$ and $r_2$ to guarantee full exploration of physically plausible values in the (p,b) plane. 
We fixed the TESS dilution factor to one based on our analysis of nearby companions. Furthermore, we added in quadrature a jitter term $\sigma_{TESS}$ to the TESS photometric uncertainties, which might be underestimated due to additional systematics in the space-based photometry. The details of the priors and  posteriors, and the description of each parameter are presented in Table~\ref{table:tessonly}. The best-fit transit model for the six sectors of TESS photometry is shown in Fig.~\ref{fig:tessall}. We also represented the phase-folded light curve in the same figure. 

\subsection{HARPS radial velocities}
\begin{figure}[h]
    \centering
    \includegraphics[width=0.47\textwidth]{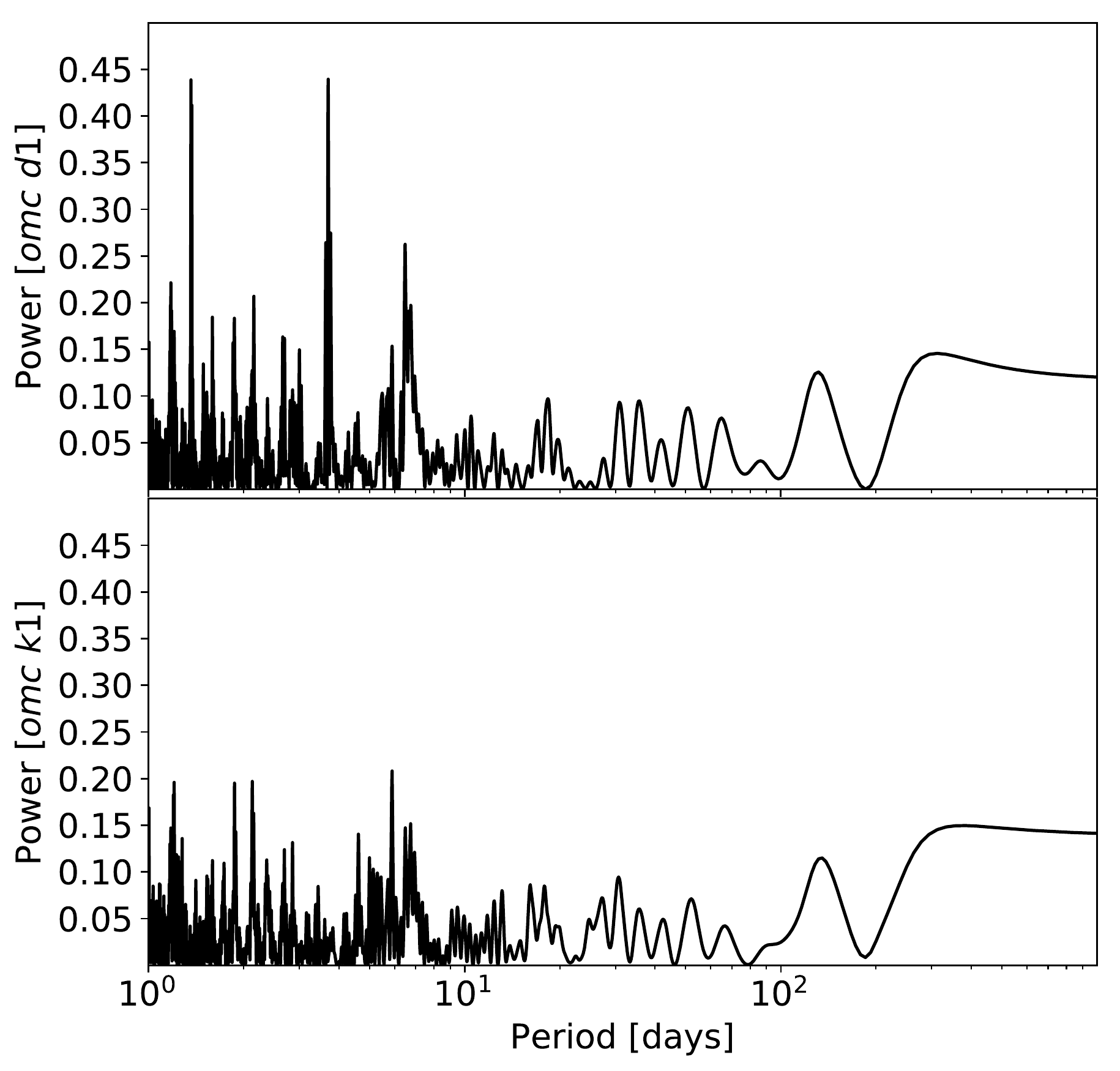}
    \caption{Analysis of the HARPS radial velocities. Periodogram for RV time series where a linear drift was subtracted is shown in the top panel. We can see a peak at 3.7~days corresponding to the period of the planet, and another at 1.36~days corresponding to its alias. Both peaks have a false alarm probability $<<1\%$. The lower panel is the periodogram for RVs where a linear drift and Keplerian (for the first planet) were subtracted; the highest peak at 5.9~days has a false alarm probability of 20.8\%.}
    \label{fig:periodoHARPS}
\end{figure}

As the star does not seem to be very active and has a long rotation period, we decided that Gaussian processes were not required to model the HARPS data. However, a small long-period trend seems visible in our dataset so we decided to add a linear drift to the fit. We first computed the periodogram of the RV time series where we subtracted the possible linear trend, and then the Keplerian corresponding to the TOI-269~b period (see Fig.~\ref{fig:periodoHARPS}). We then analyzed the radial-velocity data from HARPS using \juliet. The intercept of the linear model, which represents here the systemic velocity of the star, is computed at time $t_{\gamma} = 2\,458\,639$~BJD$_{\rm UTC}$, close to the middle of our observations. This configuration of one planet with a possible additional long-term trend (that is compatible with zero within 2$\sigma$)  in the data appears to be the simplest and model that best explains the current dataset. The priors are taken from the TESS analysis and an overview of the HARPS data. The details of the priors and  posteriors, and the description of each parameter are presented in Table~\ref{table:harpsonly}. \\

\subsection{Joint fit of all data}
\begin{figure*}[h]
    \centering
    \includegraphics[width=1.0\textwidth]{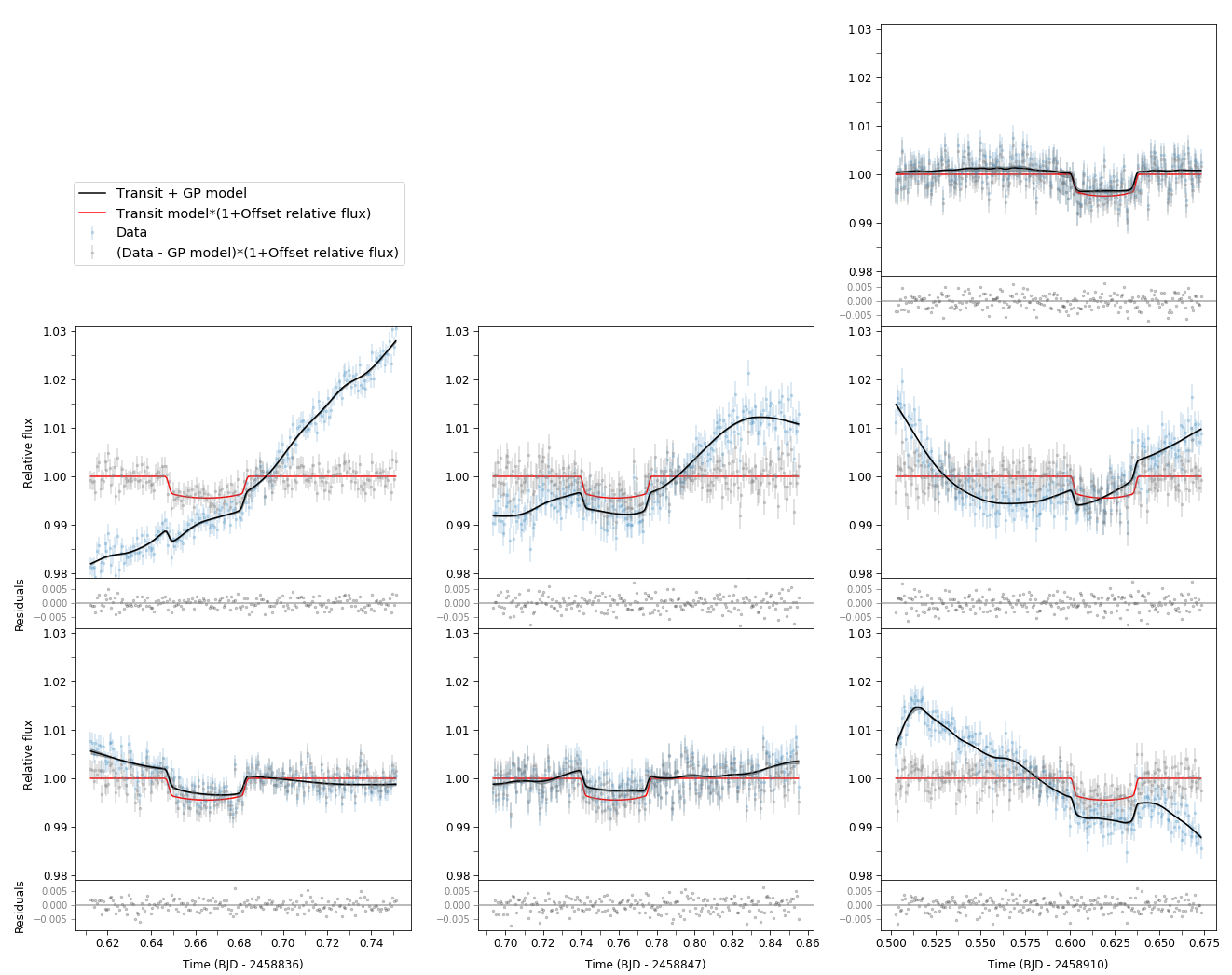}
    \caption{ExTrA light curves for the three nights of observations (columns) and for each telescope (first line corresponding to telescope 1, second line to telescope 2, and third line to telescope 3). Our raw data are shown in blue, and modeled using Gaussian processes in black. The detrended light curves are shown (gray dots), with the maximum a posteriori model transit (in red). The residuals below correspond to the modeled transit subtracted from detrended data points.}
    \label{fig:extraLC}
\end{figure*}

\begin{figure*}[h]
    \centering
    \includegraphics[width=1.0\textwidth]{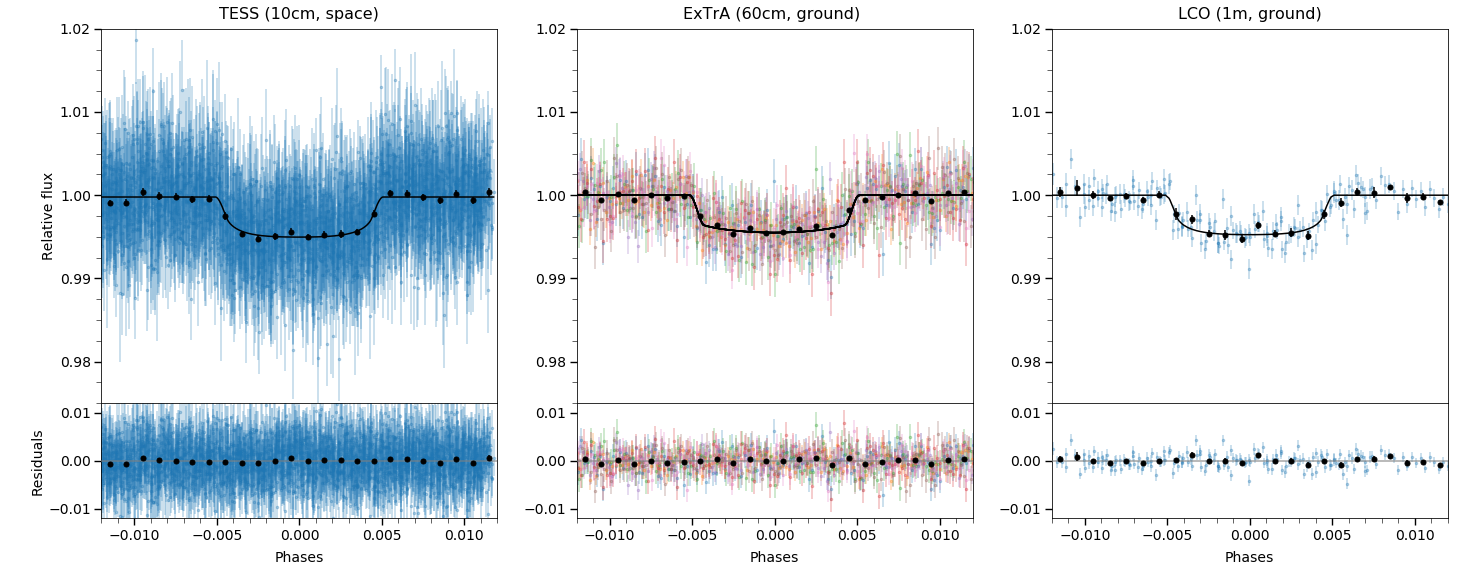}
    \caption{Photometry data for TESS (35 transits), ExTrA (7 transits; corrected for the GP component), and LCO-CTIO (2 transits; corrected for the GP component) phase-folded to the period of the planet. The black line is the maximum a posteriori model; the black points are $\sim$~5 min binned data.}
    \label{fig:allphoto}
\end{figure*}

\begin{figure*}[h]
    \centering
    \includegraphics[width=1.0\textwidth]{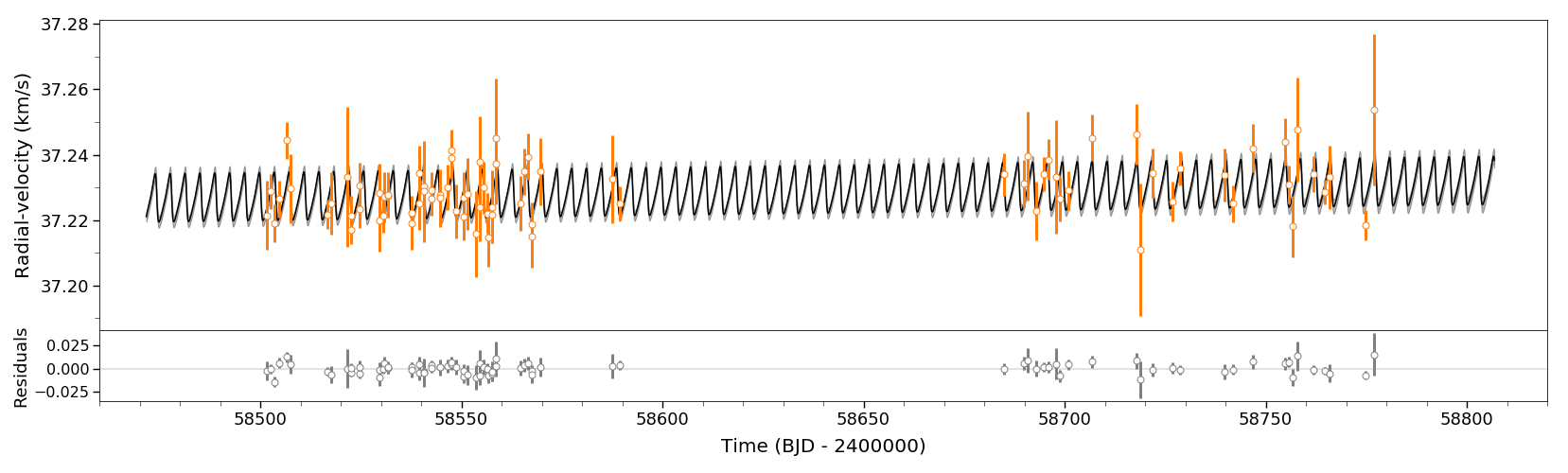}
    \caption{HARPS radial-velocity measurements for TOI-269.  The data points with errors are shown in orange; the black line and gray band correspond respectively to the median and $1\sigma$ of 1000 randomly chosen posterior samples from the joint fit.}
    \label{fig:harps}
\end{figure*}

\begin{figure}[h]
    \centering
    \includegraphics[width=0.5\textwidth]{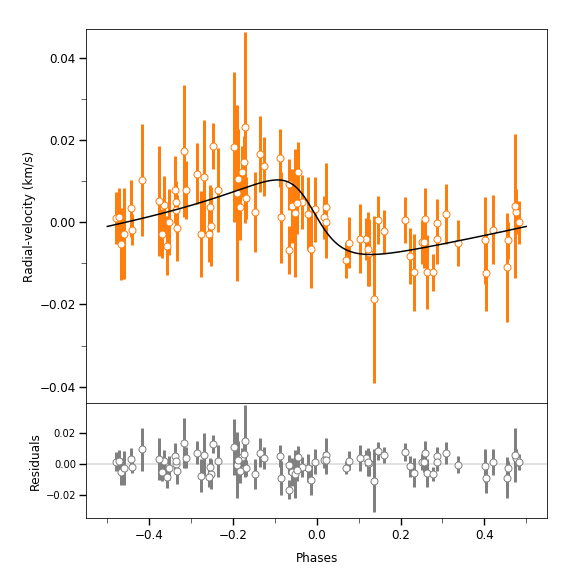}
    \caption{HARPS RVs phase-folded to the period of the transiting planet. The black line is the best-fit model from the joint fit.}
    \label{fig:harpsph}
\end{figure}

To obtain the most precise parameters of the TOI-269 system, we performed a joint analysis of the TESS, LCO, and ExTrA photometry and velocity data from HARPS using \juliet. This time we fitted for the eccentricity and applied the classical parameterization of (e,$\omega$) into ($\sqrt{e}\sin{\omega}$,$\sqrt{e}\cos{\omega}$), always ensuring that $e\leq1$. We decided to use time-dependent Gaussian processes on the ExTrA and LCO-CTIO photometry as the data had not been detrended yet. We decided on the approximate Matern kernel introduced in \cite{foreman2017} because when looking at the light curves there are no evident quasi-periodic oscillations. We only use the three-transit duration around transit data for TESS for reasons of computation time.\\
In total, we looked for the posterior distribution of 51 free parameters, and as it is a large number of parameters, we used \dynesty to perform this fit. Before the final run for the joint modeling, we constrained our priors using the previous results from \juliet in order to optimize the search for the posteriors and the computation time due to the large parameter space. Since we used rather large priors for the planetary parameters for the fit of TESS data and HARPS data separately, nested sampling is an efficient method for the exploration of the parameter space, and as most of the planetary parameters are specific to a given data type, they would not change significantly in a joint fit. Using  prior knowledge from the previous fit for the joint modeling was therefore justified. Table~\ref{table:priorjoint} and Tables~\ref{table:postparam} and ~\ref{table:postinstru} show respectively the priors and posteriors of all the parameters, along with the transit, RVs and physical parameters derived using the stellar parameters presented in Section 2.
Figure~\ref{fig:extraLC} shows the results of our joint fit to the data for ExTrA\footnote{The raw and detrended fluxes are available in electronic form at the CDS.}. The GPs we used to account for the systematics clearly capture  the observed trends in the ExTrA photometry. We show the phased transits of the photometry instruments after subtracting the GP component from the ExTrA and LCO-CTIO data in Fig.~\ref{fig:allphoto}. In Fig.~\ref{fig:harps} and  Fig.~\ref{fig:harpsph}, we present the radial-velocity component of the joint fit modeled by the median and one-sigma error of 1000 randomly chosen posterior samples, and the phase-folded data with the maximum a posteriori model. \\
This analysis revealed an eccentric orbit for TOI-269~b, which is discussed in the next section.\\

\begin{figure}[h]
    \centering
    \includegraphics[width=0.45\textwidth]{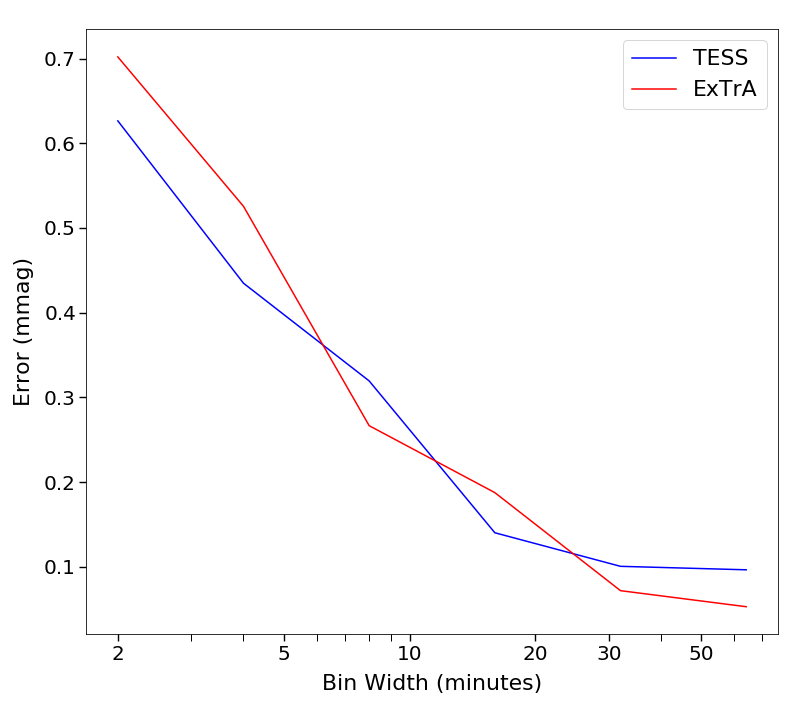}
    \caption{Dispersion of the residuals for TESS (35 combined transits) and ExTrA (7 combined transits) data for different bin sizes.}
    \label{fig:bins}
\end{figure}

In order to estimate the precision of the new instrument ExTrA, we computed the standard deviation of the residuals of TESS and ExTrA data phased at the period of the planet (see Fig.~\ref{fig:allphoto} for the residuals of the phase-folded light curves) for different bin sizes (from 2 to 64 min bins) and present the results in Fig.~\ref{fig:bins}. We can see that combining 7 ExTrA transits is similar to combining 35 TESS transits to obtain the same precision on the dispersion of the residuals, potentially correlated to the precision on the radius ratio. We established that we need to fit the TESS and ExTrA light curves simultaneously to constrain the parameters of the Gaussian processes, so that they filter ExTrA photometry on a different timescale than the transit itself, but we do obtain a better precision (1 to 5 transits) to measure the depth of the transit with fewer ExTrA observations.


\begin{table*}[h]
      \caption[]{Posterior stellar and planetary parameters obtained from our joint photometric and radial-velocity \juliet analysis for TOI-269. The posterior estimate corresponds to the median value. Error bars denote the 68\% posterior credibility intervals.}
         \label{table:postparam}
         \begin{tabular}{p{0.4\linewidth}cc}
            \hline
            \noalign{\smallskip}
            Parameter name & \text{Posterior estimate} & \text{Description}\\
            \noalign{\smallskip}
            \hline
            \noalign{\smallskip}
            Posterior parameters for TOI-269  \\
            $\rho_{\star}$ (g/cm$^3$) & $8.98^{+0.83}_{-0.81}$  & \text{Stellar density}\\
            \noalign{\smallskip}
            $q_{1,TESS}$ & $0.39^{+0.33}_{-0.24}$ & \text{Quadratic limb-darkening parameterization}\\
            \noalign{\smallskip}
            $q_{2,TESS}$ & $0.22^{+0.30}_{-0.16}$ & \text{Quadratic limb-darkening parameterization}\\
            \noalign{\smallskip}
            $q_{1,LCO}$ & $0.35^{+0.28}_{-0.19}$ & \text{Quadratic limb-darkening parameterization}\\
            \noalign{\smallskip}
            $q_{2,LCO}$ & $0.36^{+0.35}_{-0.25}$ & \text{Quadratic limb-darkening parameterization}\\
            \noalign{\smallskip}
            $q_{1,ExTrA}$ & $0.133^{+0.16}_{-0.089}$ & \text{Quadratic limb-darkening parameterization}\\
            \noalign{\smallskip}
            $q_{2,ExTrA}$ & $0.43^{+0.36}_{-0.30}$ & \text{Quadratic limb-darkening parameterization}\\
            \noalign{\smallskip}
            \hline
            \noalign{\smallskip}
            Posterior parameters for TOI-269~b \\
            $P_b$ (days) & $3.6977104^{+0.0000037}_{-0.0000037}$ & \text{Period of the planet}\\
            \noalign{\smallskip}
            $t_0$ (BJD TBD) & $2458381.84668^{+0.00033}_{-0.00033}$ & \text{Time of transit-center for the planet}\\
            \noalign{\smallskip}
            $r_{1,b}$ & $0.700^{+0.078}_{-0.15}$ & \text{Parameterization for p and b} \\
            \noalign{\smallskip}
            $r_{2,b}$ & $0.0638^{+0.0020}_{-0.0021}$ & \text{Parameterization for p and b} \\
            \noalign{\smallskip}
            $S_{1,b}=\sqrt{e_b}sin(\omega_b)$ & $0.614^{+0.077}_{-0.10}$ & \text{Parameterization for $e$ and $\omega$}\\
            \noalign{\smallskip}
            $S_{2,b}=\sqrt{e_b}cos(\omega_b)$ & $0.18^{+0.15}_{-0.16}$ & \text{Parameterization for $e$ and $\omega$}\\
            \noalign{\smallskip}
            $K_b$ (m/s) & $7.6^{+1.2}_{-1.2}$ & \text{Radial-velocity semi-amplitude of the planet}\\
            \noalign{\smallskip}
            Long-term trend in RVs\\
            \noalign{\smallskip}
            RV slope (m/s/day) & $0.0161^{+0.0081}_{-0.0081}$ & \text{Linear trend : first parameter}\\
            \noalign{\smallskip}
            RV intercept (km/s) at $t_{\gamma}$ & $37.22864^{+0.00082}_{-0.00082}$ & \text{Linear trend : second parameter}\\
            \noalign{\smallskip}
            \noalign{\smallskip}
            Derived transit and RVs parameters      \\
            $p=R_p/R_{\star}$ & $0.0638^{+0.0020}_{-0.0021}$ & \text{Planet-to-star radius ratio}\\
            \noalign{\smallskip}
            $b=(a/R_{\star})cos(i_p)$ & $0.55^{+0.12}_{-0.22}$ & \text{Impact parameter of the orbit}\\
            \noalign{\smallskip}
            $i_p$ (deg) & $88.14^{+0.78}_{-0.90}$ & \text{Inclination of the orbit}\\
            \noalign{\smallskip}
            $e_b$ & $0.425^{+0.082}_{-0.086}$ & \text{Eccentricity of the orbit}\\
            \noalign{\smallskip}
            $\omega_b$ (deg) & $74^{+15}_{-15}$ & \text{Argument of periastron}\\
            \noalign{\smallskip}
            \noalign{\smallskip}
            Derived physical parameters$^\ast$ \\
            $M_p (M_{\oplus})$ & $8.8^{+1.4}_{-1.4}$ & \text{Planetary mass (in Earth mass)}\\
            \noalign{\smallskip}
            $R_p (R_{\oplus})$ & $2.77^{+0.12}_{-0.12}$ & \text{Planetary radius (in Earth radius)}\\
            \noalign{\smallskip}
            $\rho_p (g/cm^{3})$ & $2.28^{+0.48}_{-0.42}$ & \text{Planetary density}\\
            \noalign{\smallskip}
            $a_p (au)$ & $0.0345^{+0.0015}_{-0.0015}$ & \text{Semi-major axis}\\
            \noalign{\smallskip}
            $T_{eq} (K)$ & $531^{+25}_{-25}$ & \text{Equilibrium temperature of the planet$^\dagger$}\\
            \noalign{\smallskip}
            $S (S_{\oplus})$ & $19.0^{+4.0}_{-3.3}$ & \text{Planetary insolation (in Earth flux)}\\
            \noalign{\smallskip}
            \hline
         \end{tabular}
         \begin{tablenotes}
         \small
         \item $^\ast$We sample from a normal distribution for the stellar mass, stellar radius, and stellar temperature that is based on the results from Section 2.
         \item $^\dagger$Equilibrium temperature was calculated assuming 0.3 Bond albedo and the semi-major axis distance.
         \end{tablenotes}
\label{tab:1}
\end{table*}

\subsection{TTVs analysis}\label{section.ttvs}
\begin{figure*}[h]
    \centering
    \includegraphics[width=1.0\textwidth]{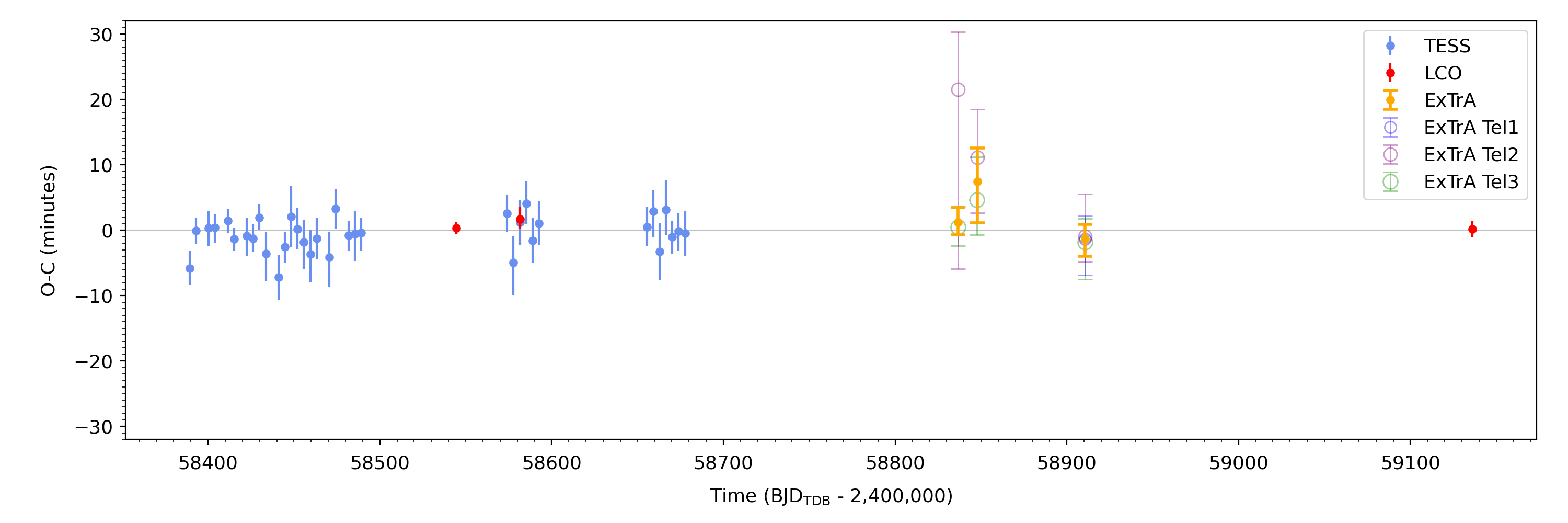}
    \caption{Transit time variations. The observation minus the calculated linear ephemeris (O-C) diagram derived using TESS (light blue), LCO-CTIO (red), and ExTrA data (dark blue corresponding to telescope 1, purple to telescope 2, green to telescope 3, and yellow to a transit time fit of all the telescopes available for each night).}\label{fig:ttvs}
\end{figure*}

We calculated the transit timing variations (TTVs) with all the photometric datasets (TESS, LCOGT, ExTrA) using \juliet. Instead of fitting a period $P$ and a time-of-transit center $t_0$, \juliet looks for the individual transit times. We fitted each ExTrA transit individually and one transit time for each night of observations, which is more coherent for the analysis. Combining the three telescopes reduces the error bars on each ExTrA estimation. The results of the analysis showing the difference between the observed transit times and the calculated linear ephemeris from all the transits is presented in Fig.~\ref{fig:ttvs}. No significant variation appears in the data. \\

\section{Discussion}\label{section.discussion}

\subsection{Mass-radius diagram and internal structure}
\begin{figure}[h]
    \centering
    \includegraphics[width=0.49\textwidth]{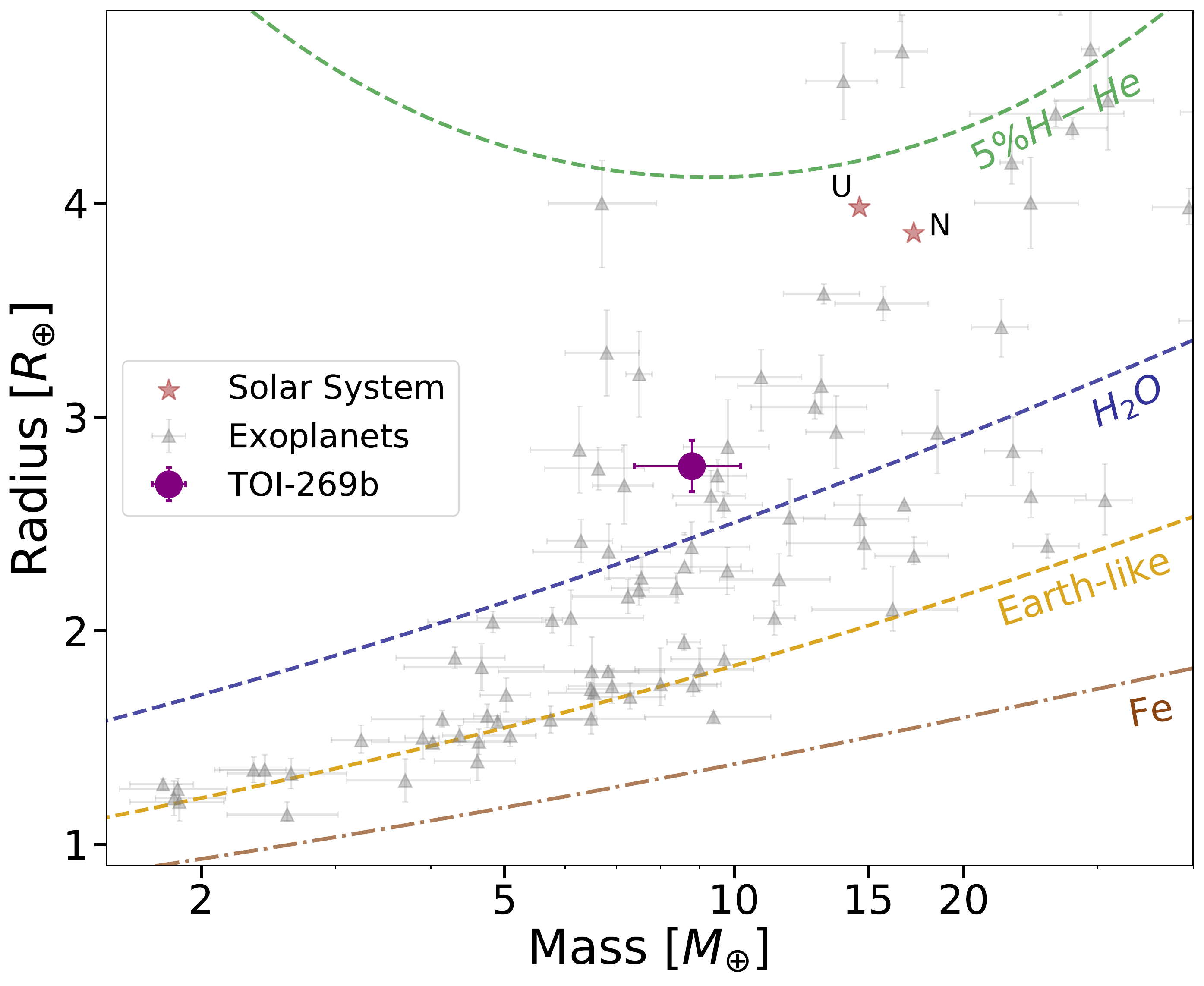}
    \caption{Mass-radius diagram of exoplanets with accurate mass and radius determination \cite[][]{otegi2020}. Also shown are the composition lines of iron, an Earth-like planet, pure water, and 95\% water+5\% H-He subjected to an insolation flux of 20$S_{\oplus}$ (similar to that of TOI-269b). The EOSs used for the iron core are taken from \cite{Hakim2018}, the EOS of the silicate-mantle is calculated with {{\sc \tt PERPLE\_X}\xspace} from \cite{Connolly09}, and for  pure-water the AQUA EOS from \cite{Haldemann2020} is used. }
    \label{fig:massrad}
\end{figure}

TOI-269~b, with a radius of $2.77\pm0.12~R_{\oplus}$ and a mass of $8.8\pm1.4~M_{\oplus}$, lies within the population of sub-Neptunes. With a stellar irradiation of 19 times the Earth, TOI-269~b lies in the gas-dominated sub-Neptune population described by \cite{fulton2018}. Figure~\ref{fig:massrad} shows M-R curves tracing the compositions of pure-iron, Earth-like, and pure-water planets and a planet with 95\% water and 5\% H-He atmosphere subjected to a stellar radiation of $F/F_{\oplus}$= 20 (similar to that of TOI-269~b). For reference, also shown are exoplanets with accurate and reliable mass and radius determinations \cite[][accessible on the Data \& Analysis Center for Exoplanet,  DACE\footnote{https://dace.unige.ch/exoplanets/}]{otegi2020}. TOI-269~b sits above the pure-water curve and below the 5\% H-He curve, implying that the H-He mass fraction is unlikely to exceed a few percent. \\

Determining the planetary internal structure is extremely challenging since various  compositions can lead to identical mass and radius (e.g., \citealt{Rogers10}, \citealt{Lopez14}, \citealt{Dorn15,Dorn17}, \citealt{Lozovsky18}, and \citealt{otegi-20-2}). With the aim of at least partially breaking the degeneracy, and in order to determine how well a given interior model compares with the other possible models that also fit the data, we use a generalized Bayesian inference method with a Nested sampling scheme. This approach allows us to quantify the degeneracy and correlation of the planetary structural parameters and to estimate the most likely region in the parameter space. We modeled the interior of TOI-269~b considering a pure-iron core, a silicate mantle, a pure-water layer, and a H-He atmosphere. The equations of state (EOSs) used for the iron core are taken from \cite{Hakim2018},  the EOS of the silicate-mantle is calculated with {{\sc \tt PERPLE\_X}\xspace} from \cite{Connolly09} using the thermodynamic data of \cite{stixrude_thermodynamics_2011}, and the EOS for the H-He envelope are from \cite{Chabrier2019} assuming a proto-solar composition.  For  the pure-water layer the AQUA EOS from \cite{Haldemann2020} is used.  The thickness of the planetary layers were set by defining their masses and solving the structure equations. To obtain the transit radius, we follow \cite{Guillot2010} and evaluate the location where the chord optical depth $\tau _{ch}$ is $2/3$.  It has been suggested that stellar abundances can be used as a proxy for the planetary bulk abundances to reduce the degeneracy (\citealt{Dorn-17}, \citealt{Brugger-17}, and \citealt{otegi-20-2}), but the result is debated \cite[][]{Plotnykov}, and we therefore opted not to use them.

\begin{figure*}[h]
    \centering
    \includegraphics[width=1.0\textwidth]{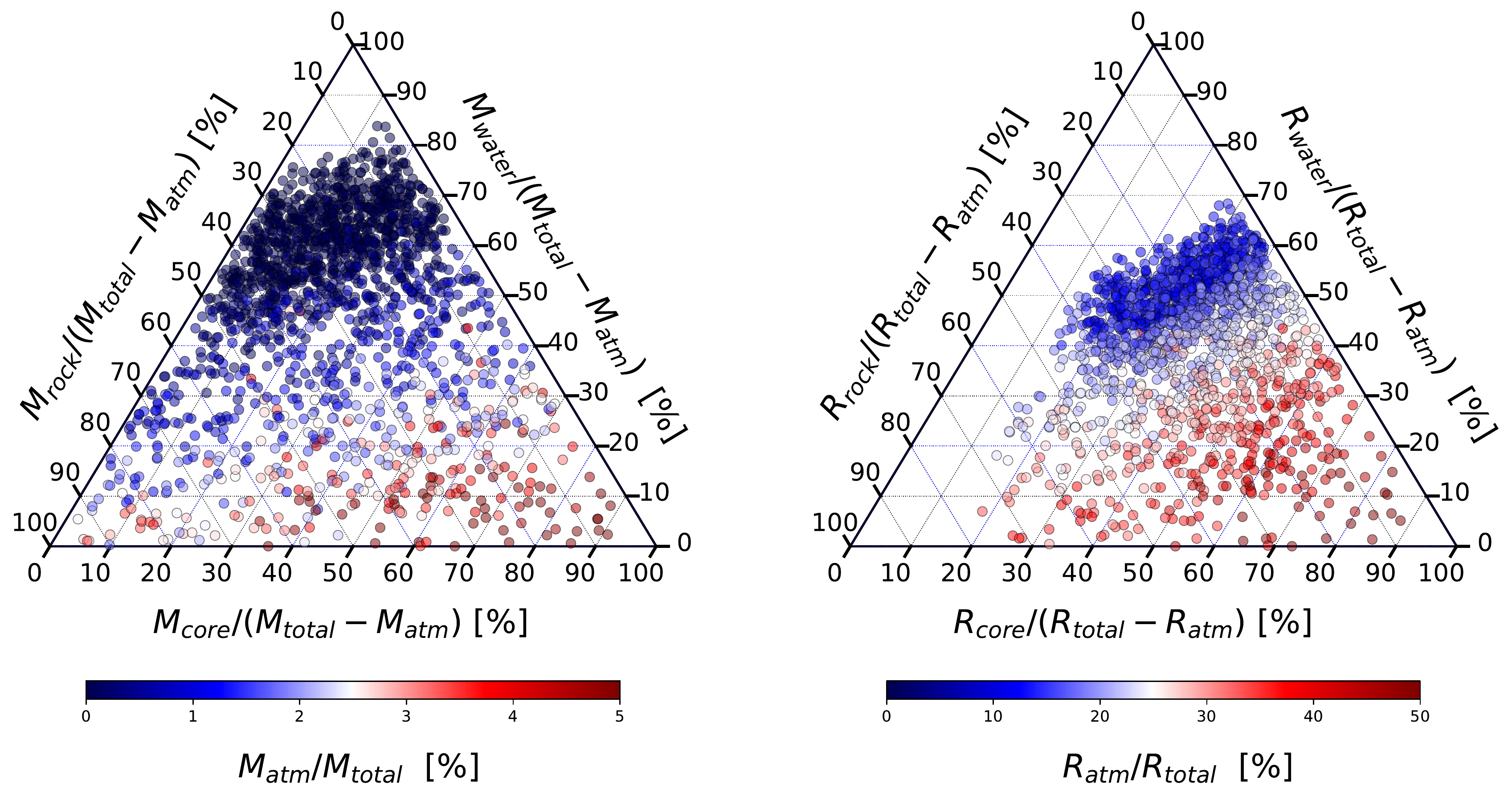}
    \caption{Ternary diagram of the inferred internal composition of TOI-269~b in mass (left) and radius (right).  }
    \label{fig:composition}
\end{figure*}

Figure~\ref{fig:composition} shows a ternary diagram and an illustration of the inferred internal composition of TOI-269~b. The ternary diagram shows the strong degeneracy of internal composition leading to the same mass and radius, which cover almost the whole diagram. We find a median H-He mass fraction of 0.8\%, which corresponds to a lower bound since enriched H-He atmospheres are more compressed and, therefore increase the planetary H-He mass fraction. Formation models suggest that sub-Neptunes are very likely to be formed via envelope enrichment \cite[][]{Venturini2017}. We also find that TOI-269~b can have a significant water layer, accounting for nearly half of the planetary mass with a thickness of about one Earth radius ($1.02 ^{+0.39} _{-0.37}$). The iron core and silicate mantle have estimated relative mass fractions of 19\% and 26\%, with large uncertainties. The degeneracy between the core and mantle in this M-R regime does not allow us to accurately estimate the masses of these two constituents. Since interior models cannot distinguish between water and H-He as the source of low-density material, we also ran a three-layer model which leaves out the H$_{2}$O envelope. Under this assumption we find that the planet would be nearly $4^{+1}_{-1}\%$ H-He, $43^{+24}_{-23}\%$ iron, and $52^{+22}_{-23}\%$ rock by mass. These estimations with the three-layer model set maximum limits since any water added would decrease these mass fractions. 

\subsection{A significantly eccentric orbit}

\begin{figure}[h]
    \centering
    \includegraphics[width=0.5\textwidth]{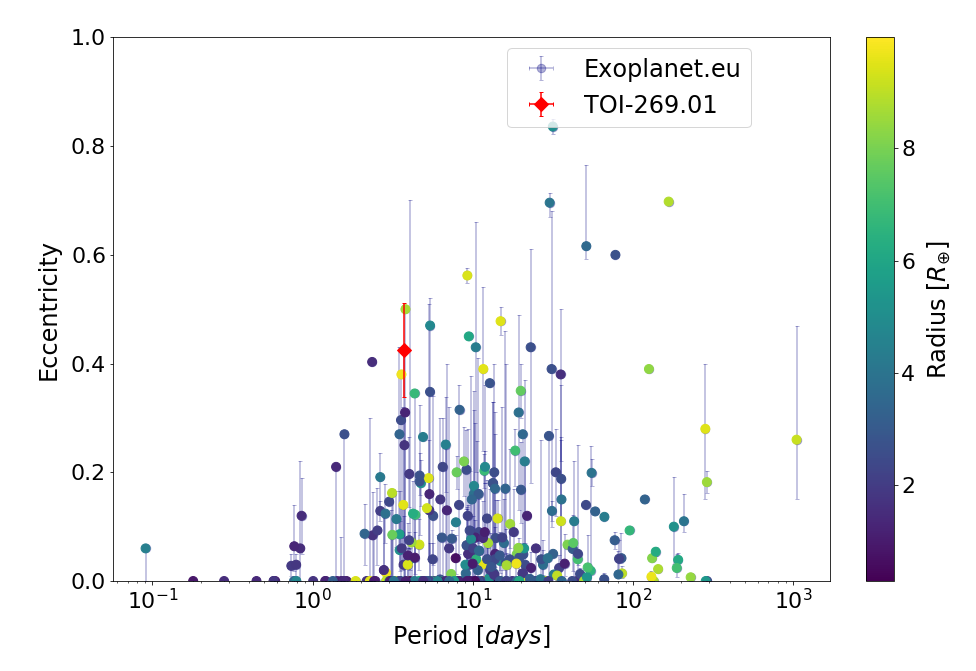}
    \caption{Orbital eccentricity as a function of the orbital period with the planetary radius (see  color scale at right) from the Exoplanet.eu catalog.}\label{fig:perecrad}
\end{figure}

In our joint fit of photometric and velocimetric data, we derived a non-zero eccentricity of $0.425^{+0.082}_{-0.086}$, with 99\% of the posterior values between 0.210 and 0.577. As shown on an eccentricity-period diagram with exoplanets smaller than 10 Earth radii (Fig.~\ref{fig:perecrad}), TOI-269~b has a remarkable eccentricity with almost the highest value for planets with periods shorter than 10 days.

\begin{figure}[h]
    \centering
    \includegraphics[width=0.5\textwidth]{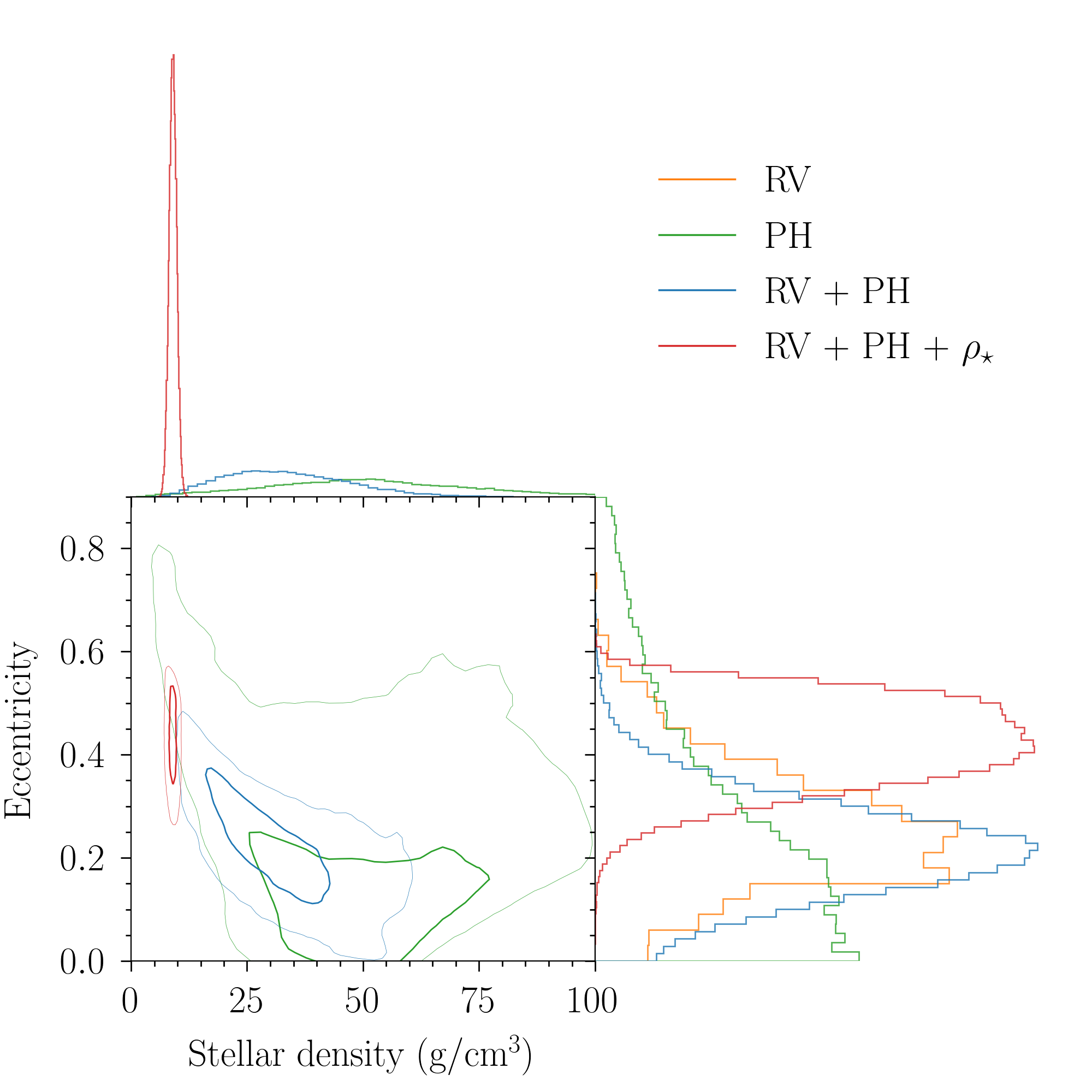}
    \caption{Two-parameter joint posterior distributions for the stellar density and planet eccentricity for three different analyses (photometry only; joint radial velocity and photometry; and joint radial velocity and photometry with a stellar density prior computed from the stellar radius and mass derived in Section~\ref{section:stellarparameters}). The 39.3\% and 86.5\% two-variable joint confidence regions are denoted by a thick and a thin line, respectively. Histograms of the marginal posterior samples are presented at the top and at the right (for the eccentricity, the radial velocity-only  analysis is also shown).}\label{fig:corner}
\end{figure}

To understand which part of our data constrains the eccentricity most, we applied different fits and compared their posterior distributions (Fig.~\ref{fig:corner}). First, we fitted only the photometry using data from TESS, ExTrA, and LCO, and without a prior on the stellar density. This yielded a broad distribution of eccentricities showing that, although compatible with zero, the photometry alone poorly constrains the eccentricity value. Second, we fitted only the HARPS radial velocities, and we obtained an eccentricity of $0.25^{+0.13}_{-0.10}$, close to the 2$\sigma$ threshold chosen by \citet{lucy1971} to consider an eccentricity value to be significantly different than zero. As expected, the posterior distribution is also narrower, showing that radial velocity alone constrains the eccentricity more than the photometry alone does. Actually, when fitting both photometry and radial-velocity data together (but still without a prior on the stellar density), we obtained a similar posterior distribution on eccentricities than that of our fit with only radial velocity.

In Fig.~\ref{fig:corner} we also show the posterior distributions for the stellar density when relevant (i.e., not when fitting RV alone). With posterior values ranging from 10 to 100 $g/cm^3$, we see that the stellar density is poorly constrained by either photometry alone or both photometry and radial velocity. Most of these values are actually too high for such an early M dwarf. We also see that there is a strong correlation between the stellar density and the eccentricity of the planet. This introduces the importance of the prior on the stellar density. In Fig.~\ref{fig:corner}, we add the result of our joint fit of both photometric and velocimetric data {including} the prior knowledge on the stellar density (see Section~\ref{section:stellarparameters}). The narrow prior on stellar density selects a fraction of an otherwise broad $e-\rho_{star}$ posterior density distribution. At this point we can thus conclude that if some of the eccentricity is already encoded in the radial-velocity data, most of the constraints eventually come from the prior on the stellar density.

To lend more credit to this remarkable result, we also modeled the data by imposing a circular orbit to TOI-269~b. It leads to a stellar density $>25.8 g/cm^3$ with a 99\% confidence level, which is not compatible with the radius and mass estimation we provided in Section~\ref{section:stellarparameters}, and therefore also excludes that the orbit could be circular. As a side note, we also tried to account for the finite integration time of the observations, in case the too long integration times would bias the light curve toward lower stellar densities \citep{kipping2010}, but it did not change our posterior distribution for the stellar density.

Granted with a robust non-zero eccentricity, we turn to possible explanations. Because the star is likely a few billion years old (Section~\ref{section:stellarparameters}), TOI-269~b probably did not acquire its eccentricity recently. It may have reached its present orbit with planet-planet migration and acquired a high eccentricity in the process. We looked at the circularization timescale following \cite{patra2017}. Determining which tidal quality factor to use for a specific exoplanet is not simple, so we looked at the problem the other way around. Given the two ages that we estimated for TOI-269, we would need a tidal quality factor Qp larger than 1.5x$10^5$ and 2.8x$10^5$ in order to explain why the exoplanet is not on a circular orbit yet. This range of values is high compared to the tidal quality factors calculated for the Solar System planets, where Qp is around 10-100 for the rocky planets \citep{Goldreich66} and up to $\sim~30000$ for Jupiter using the excitation frequency of Io \citep{Lainey09}. However, in studies that target exoplanets (e.g.,   \citealt{Hansen12}) it was  shown that Qp can go as high as $10^7-10^8$ for exoplanets that orbit close to the host star. The discrepancy between the values of Qp for Jupiter and for the close-in exoplanets can be explained by the fact that these planets are likely to be in synchronous rotation, contrary to Jupiter, which has a fast rotation. A slower synchronous rotation would make the dissipation via inertial waves less efficient \citep[e.g.,][]{Ogilvie04}, and therefore the planets would take more time to circularize. The observations can give us an upper value for Qp, in order to explain why the planet is not circularized yet, but we would need models of the internal structure and tidal dissipation specifically of TOI-269~b to constrain this factor more closely, and helps us give a lower limit on Qp for example. It is also interesting to note that \citet{correia2020} proposed other mechanisms to explain non-null eccentricity of other warm Neptunes, including excitation from distant planet or atmospheric escape. We note that the RVs include a possible trend (0.0161$\pm$0.0081 m/s/day), and therefore a possible hint of an outer companion. Given the $\sim$~275-day time span of observations with HARPS, if the possible companion were to have an orbital period of twice this time span, its mass would be $30\pm15$ $M_{\oplus}$.


\subsection{Potential for atmospheric characterization}\label{atm_char}

\begin{figure}[h]
    \centering
    \includegraphics[width=0.5\textwidth]{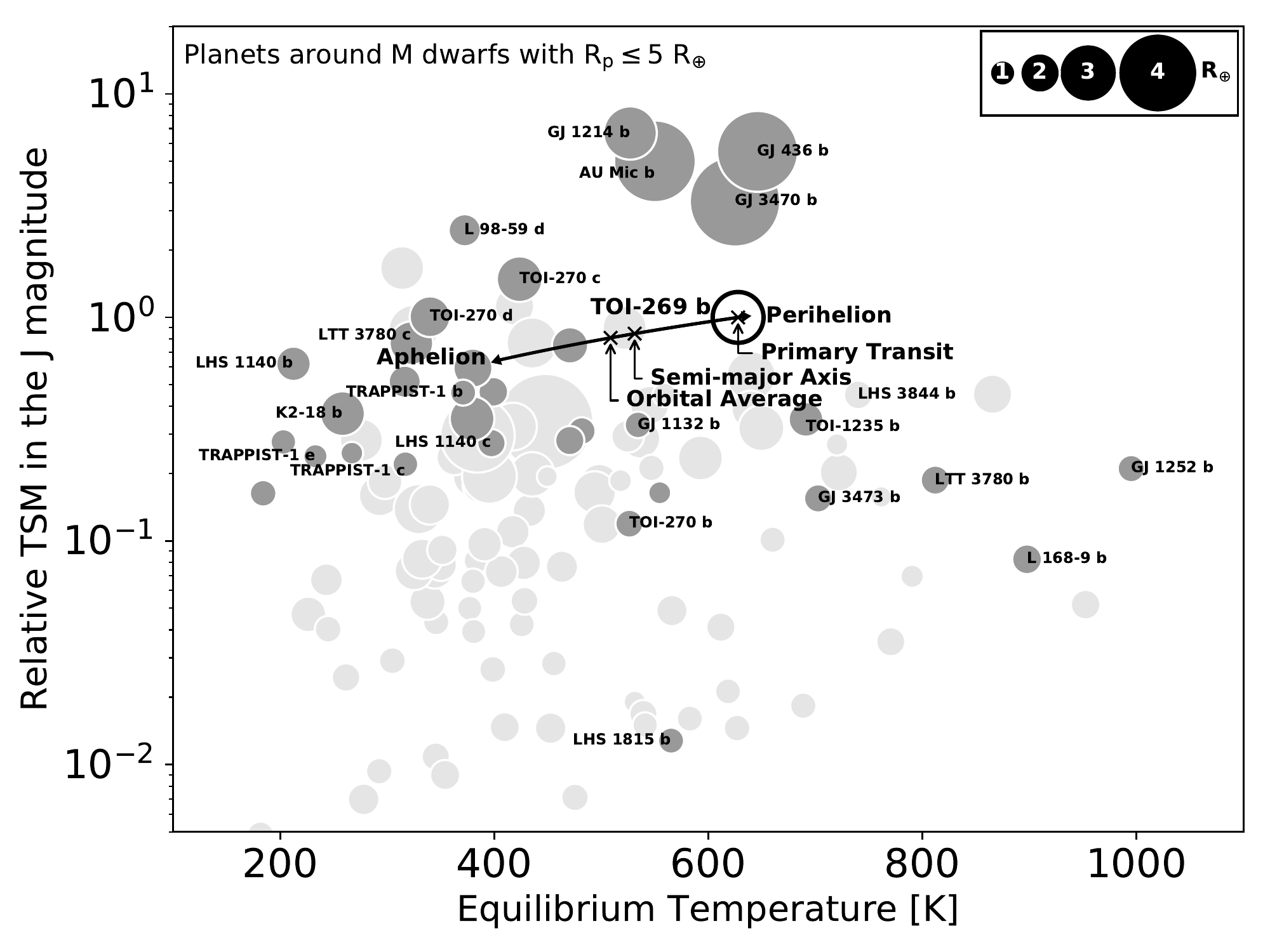}
    \caption{Transmission spectroscopic metric in the $J$ magnitude as a function of the planetary equilibrium temperature. The TSM values are shown relatively to the TOI-269~b (eccentricity-corrected) reference value of $89^{+19}_{-14}$. Only planets orbiting M dwarf (${\rm T_{\rm eff}}\leq3875$ [K]) and with ${\rm R_p}\leq 5 {\rm R_\oplus}$ were selected. TOI-269~b is shown as a black circle. The black arrow shows the range of equilibrium temperature depending on planet position during the orbit. The crosses show the locations of the orbital average, the semi-major axis, and the primary transit. Other planets with and without measured masses are shown in dark gray and light gray, respectively.}
    \label{fig:TSM_K18}
\end{figure}

We calculated the transmission spectroscopic metric (TSM) in the J magnitude using the equation from \cite{kempton2018} and obtained a value of TSM = $75^{+16}_{-12}$. This value takes into account the equilibrium temperature when the planet is at a distance corresponding to its semi-major axis. For an eccentric orbit and for transmission spectroscopy it would be more exact to take into account the planet equilibrium temperature at the exact transit moment (see details below). By doing so  a corrected TSM of $89^{+19}_{-14}$ is computed. To put this value into context, we also computed the TSM for exoplanets with ${\rm R_p}\leq 5 {\rm R_\oplus}$ orbiting M dwarf host stars (${\rm T_{\rm eff}}\leq3875$ [K]) using the NASA exoplanet archive on February 16, 2021. The TSM values are shown relatively to the TOI-269~b value in Fig.\ref{fig:TSM_K18}. 
TOI-269~b lies among the best targets of its category (sub-Neptunes around M dwarfs). So far, the atmospheric characterization of sub-Neptune planets mostly delivered non-detections \citep[e.g., on the warm GJ~1214~b by][]{Kreidberg2014}, and the few detections obtained did not bring conclusive results on their actual atmospheric contents. For example, the more temperate K2-18~b \citep{Cloutier2017} exhibits a near-infrared absorption in its transmission spectrum first interpreted as due to water (\citealt{Benneke2019}, \citealt{Tsiaras2019}), but \citet{Bezard2020} argued that methane could be responsible for this feature. Thus, it would be very interesting to probe with transmission spectroscopy the atmospheres of newly discovered temperate and warm sub-Neptunes such as TOI-269~b, TOI-270~c or~d (\citealt{Gunter2019}, \citealt{vaneylen21}), or LTT 3780~c (\citealt{cloutier2020b}, \citealt{Nowak2020}) in order to see if they exhibit atmospheric features and if so, what  their natures are. Finally, detecting any atmospheric features could help in constraining the extent and the mean molecular weight of the atmosphere, and give some constraints on possible interior structures (see Section~5.1).

Interestingly, the orbit of TOI-269~b is found to be significantly eccentric. The change in stellar irradiation due to the eccentricity throughout the orbital revolution brings a change in equilibrium temperature by a factor of $\sqrt{\frac{1+e}{1-e}}\sim1.6\pm0.1$. Thus, the planet temperature passes from $\sim400$ K near aphelion to $\sim630$ K near perihelion. Depending of the actual atmospheric conditions, some species such as sulfuric acid may condense and vaporize throughout an orbital revolution, but details of such phenomena are beyond the scope of this paper. We also note that the primary transit happens $\sim1.5$h after perihelion (at $\sim1.02$ times the perihelion radius), when the planet is the hottest. This means that TOI-269~b is a slightly better target for transmission spectroscopy than by only considering the semi-major axis or the orbital average equilibrium temperature (see Fig.\ref{fig:TSM_K18}).

Moreover, the currently found high eccentricity could hint that the planet may have recently arrived in its position. As the host star is an M dwarf with a convective envelope, strong tidal interactions are supposed to circularize the planetary orbit rapidly \citep{correia2020}. If the planet migrated recently (e.g., due to a Lidov-Kozai mechanism), it is possible that strong to moderate atmospheric escape also started recently and would be still in process \citep{Bourrier2018a}. Such a scenario is supported by the observation of atmospheric evaporation in the Neptune-sized planets GJ 436~b \citep{Ehrenreich2015}, GJ 3470~b (\citealt{Bourrier2018b}, \citealt{Palle2020}), and HAT-P-11~b \citep{Allart2018}, which  all possess eccentric orbits. Observing evaporation in an eccentric sub-Neptune would be very informative on the architecture and history of the system. For example, this is possible by detecting the meta-stable helium infrared triplet at 1\,083 nm (\citealt{Oklopcic2018}, \citealt{Spake2018}) at high resolution (\citealt{Allart2018}, \citealt{Nortmann2018}). Even though no helium detection has been reported to date for sub-Neptune planets (\citealt{Kasper2020}, \citealt{Gaidos2020a,Gaidos2020b}), TOI-269~b could have a helium signal of 0.5--1.5\% \citep[$\gtrsim$50--150 scale-height, depending on the atmospheric escape rate, thermospheric temperature, and stellar high-energy irradiation;][]{Nortmann2018,Kasper2020} that could be detected with several transit observations with a high-resolution infrared spectrograph such as NIRPS \citep{Bouchy2017}.

\section{Conclusions}
We presented the discovery and characterization of a sub-Neptune transiting the M dwarf TOI-269. The planet was detected by the TESS mission, then confirmed via ground-based transit follow-up observations with multiple instruments and from precise RV measurements with HARPS by measuring its mass.
TOI-269~b contributes to the TESS Level One Science Requirement of delivering 50 transiting small planets (with radii smaller than $4R_e$) with measured masses to the community. It will be re-observed in five sectors during the TESS year-3 extended mission from Sept 2020 to April 2021, which will provide new timing to perform a better TTV analysis.

The estimated average density of TOI-269~b is $2.28^{+0.48}_{-0.42} g/cm^{3}$, which is significantly lower than the typical density of rock planets and indicates the presence of a volatile envelope. Internal structure models with four layers (an iron core, silicate mantle, water ocean, and H-He atmosphere), although strongly degenerated, suggest that TOI-269~b has between 0.4\% and 4\% H-He by mass. Our analysis suggests that TOI-269~b would be an interesting target for atmospheric characterization in order to compare it with other sub-Neptunes.

The ExTrA observations of this target allowed us to test the precision of our instrument. Early M dwarfs with many TESS transits are not the primary targets for ExTrA. It will be more competitive around late stars with one or a few TESS transits in order to obtain a better precision for the planetary radius and for the transit timings. As for measuring radii, we were able to show here that the precision of one transit with ExTrA is comparable to that of five transits with TESS (see Fig.~\ref{fig:bins}). ExTrA could also detect other planets in already-known planetary systems.

With such a high eccentricity, TOI-269~b is reminiscent of GJ\,436~b \citep{Bourrier2018b} and  follow-up observations similar to those conducted for this planet would thus provide an instructive comparison. TOI-269~b is probably too far away to detect an atmospheric  escape in Ly-$\alpha$ \citep{Ehrenreich2015}, but could be attempted with transmission spectroscopy of the helium triplet (Sect.~\ref{atm_char}). In addition, further RV monitoring will be valuable and, actually, already anticipated with the forthcoming near-infrared spectrograph NIRPS \citep{Bouchy2017}. They may detect the companion responsible for the possible RV trend that we have identified in this paper, and will also measure the Rossiter-McLaughlin anomaly to see if, like GJ\,436~b, TOI-269~b also has a misaligned orbit \citep{Bourrier2018b}. 

\begin{acknowledgements}
We are grateful to the ESO/La Silla staff for their continuous support. We thank the referee for his careful reading of the manuscript and his thoughtful comments.
We acknowledge funding from the European Research Council under the ERC Grant Agreement n. 337591-ExTrA. 
This paper includes data collected by the TESS mission. Funding for the TESS mission is provided by the NASA Explorer Program.
We acknowledge the use of public TESS Alert data from the pipelines at the TESS Science Office and at the TESS Science Processing Operations Center. Resources supporting this work were provided by the NASA High-End Computing (HEC) program through the NASA Advanced Supercomputing (NAS) Division at Ames Research Center for the production of the SPOC data products.
This work makes use of observations from the LCOGT network.
This work made use of \texttt{tpfplotter} by J. Lillo-Box (publicly available in www.github.com/jlillo/tpfplotter), which also made use of the python packages \texttt{astropy}, \texttt{lightkurve}, \texttt{matplotlib} and \texttt{numpy}. We thank the Swiss National Science Foundation (SNSF) and the Geneva University for their continuous support to our planet search programs. This work has been in particular carried out in the frame of the National Centre for Competence in Research ‘PlanetS’ supported by the Swiss National Science Foundation (SNSF). 
We thank La Silla observatory staff for their support.
N.A-D. acknowledges the support of FONDECYT project 3180063.
A.W. acknowledges the financial support of the SNSF by grant number P400P2\_186765.
This material is based upon work supported by the National Science Foundation Graduate Research Fellowship Program under Grant No. DGE-1650115.
This work was supported by FCT - Funda\c{c}\~ao para a Ci\^encia e a Tecnologia through national funds and by FEDER through COMPETE2020 - Programa Operacional Competitividade e Internacionaliza\c{c}\~ao by these grants: UID/FIS/04434/2019; UIDB/04434/2020; UIDP/04434/2020; PTDC/FIS-AST/32113/2017 \& POCI-01-0145-FEDER-032113; PTDC/FIS-AST/28953/2017 \& POCI-01-0145-FEDER-028953.
Based in part on observations obtained at the Southern Astrophysical Research (SOAR) telescope, which is a joint project of the Minist\'{e}rio da Ci\^{e}ncia, Tecnologia e Inova\c{c}\~{o}es (MCTI/LNA) do Brasil, the US National Science Foundation’s NOIRLab, the University of North Carolina at Chapel Hill (UNC), and Michigan State University (MSU).
FPE would like to acknowledge the Swiss National Science Foundation (SNSF) for supporting research with HARPS through the SNSF grants nr. 140649, 152721, 166227 and 184618.
J.R.M acknowledges continuous grants from CNPq, CAPES and FAPERN brazilian agencies. C.D. acknowledges support from the Swiss National Science Foundation under grant PZ00P2\_174028.
This work is supported by the French National Research Agency in the framework of the Investissements d’Avenir program (ANR-15-IDEX-02), through the funding of the "Origin of Life" project of the Univ. Grenoble-Alpes.
B.L.C.M and I.C.L. acknowledge continuous grants from CNPq, CAPES and FAPERN brazilian agencies.
We thank Emeline Bolmont for interesting discussion about tidal dissipation.
\end{acknowledgements}

\bibliographystyle{aa}
\nocite{*} 
\bibliography{refs.bib}

\begin{appendix} 

\section{ExTrA field of view of TOI-269}\label{section:FoV}

\begin{figure*}[h]
   \centering
   \includegraphics[width=0.9\textwidth]{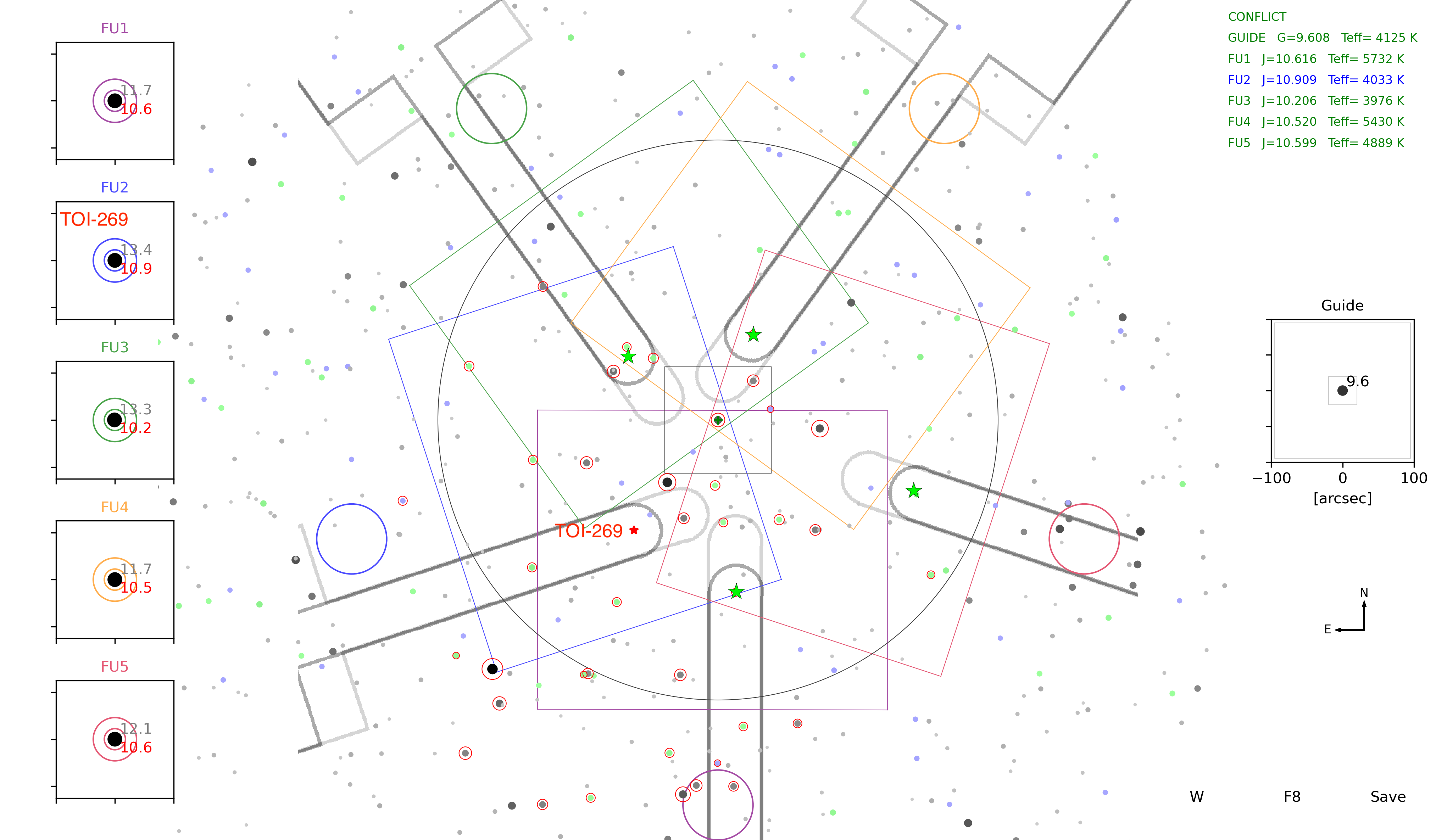}
      \caption{Snapshot of the software for preparing observations for ExTrA. The field of view of one telescope has a diameter of 1$^{\rm o}$ (gray circle). Each of the five field units (FUs) can be positioned in the part of the focal plane shown with a rectangle of corresponding color.
      In this figure the field of TOI-269 is shown. Each FU is placed over a particular comparison star (green stars) or the target (red star).
      On the left the location of each FU is magnified and the fiber apertures are drawn to scale (colored circles; the first circle is 4", the second circle is 8"). This is used to  check if other stars appear near the fibers. In red is the J magnitude of the star, in gray the Gaia magnitude. 
      The guiding star is located at the center and a magnified image is shown on the right. 
      The five arms correspond to the selected stars with their 2MASS J magnitude and Gaia DR2 temperature indicated at the top right of the figure.}
    \label{fig:field}
\end{figure*}

\clearpage

\section{Prior and posterior values for the different fits}\label{section:params}

\begin{table*}[t]
      \caption[]{Prior and posterior values for the photometry fit of TESS data. The posterior estimate corresponds to the median value. Error bars denote the 68\% posterior credibility intervals.}\label{table:tessonly}
         \label{}
         \begin{tabular}{p{0.25\linewidth}ccc}
            \hline
            \noalign{\smallskip}
            Parameter Name & \text{Prior} & \text{Posterior} & \text{Description} \\
            \noalign{\smallskip}
            \hline
            \noalign{\smallskip}
            Parameters for TOI-269~b      \\
            $P_b$ (days) & $\mathcal{N}(3.7,0.1^2)$ & $3.697722^{+0.000017}_{-0.000016}$ & \text{Period of the planet} \\
            \noalign{\smallskip}
            $t_0$ (BJD TDB) & $\mathcal{N}(2458381.84,0.1^2)$ & $245831.84601^{+0.00058}_{-0.00058}$ & \text{Time of transit-center for the planet} \\
            \noalign{\smallskip}
            $\rho_{\star}$ ($g/cm^3$) & $\mathcal{N}(8.760,0.870)$ & $9.10^{+0.82}_{-0.74}$ & \text{Stellar density}  \\
            \noalign{\smallskip}
            $r_{1,b}$ & $\mathcal{U}(0,1)$+ & $0.8983^{+0.0091}_{-0.0099}$ & \text{Parametrization for p and b} \\
            \noalign{\smallskip}
            $r_{2,b}$ & $\mathcal{U}(0,1)$ & $0.0725^{+0.0024}_{-0.0021}$ & \text{Parametrization for p and b} \\
            \noalign{\smallskip}
            $e_b$ & 0. & 0. & \text{Eccentricity} \\
            $\omega_b$ (degrees) & 90. & 90. & \text{Argument of periastron} \\
            \noalign{\smallskip}
            \hline
            \noalign{\smallskip}
            Parameters for TESS \\
            $D_{TESS}$ & 1. & 1. & \text{Dilution factor for TESS} \\
            \noalign{\smallskip}
            $M_{TESS}$ & $\mathcal{N}(0,0.1^2)$ & $-0.000015^{+0.000016}_{-0.000016}$ & \text{Relative flux offset for TESS} \\
            \noalign{\smallskip}
            $\sigma_{w,TESS}$ (ppm) & log$\mathcal{U}(0.01,100)$ & $0.58^{+8.1}_{-0.55}$ & \text{Extra jitter term for TESS lightcurve} \\
            \noalign{\smallskip}
            $q_{1,TESS}$ & $\mathcal{U}(0,1)$ & $0.128^{+0.22}_{-0.099}$ & \text{Quadratic limb-darkening parametrization} \\
            \noalign{\smallskip}
            $q_{2,TESS}$ & $\mathcal{U}(0,1)$ & $0.42^{+0.34}_{-0.28}$ & \text{Quadratic limb-darkening parametrization} \\
            \noalign{\smallskip}
            \hline
         \end{tabular}
\label{tab:1}
\end{table*}

\begin{table*}[t]
      \caption[]{Prior and posterior values for the RVs fit of the HARPS measurements. The posterior estimate corresponds to the median value. Error bars denote the 68\% posterior credibility intervals.}\label{table:harpsonly}
         \label{}
         \begin{tabular}{p{0.25\linewidth}ccc}
            \hline
            \noalign{\smallskip}
            Parameter Name & \text{Prior} & \text{Posterior} & \text{Description} \\
            \noalign{\smallskip}
            \hline
            \noalign{\smallskip}
            Parameters for TOI-269~b      \\
            $P_b$ (days) & $\mathcal{N}(3.697722,0.000017^2)$ & $3.697723^{+0.000018}_{-0.000017}$ & \text{Period of the planet} \\
            \noalign{\smallskip}
            $t_0$ (BJD TDB) & $\mathcal{U}(2458381.84601,0.00058)$ & $3.697722^{+0.000017}_{-0.000016}$ & \text{Time of transit-center for the planet} \\
            \noalign{\smallskip}
            $K_b$ (m/s)& $\mathcal{U}(0,20)$ & $7.2^{+1.3}_{-1.2}$ & \text{Radial-velocity semi-amplitude of the planet} \\
            \noalign{\smallskip}
            $S_{1,b}=\sqrt{e_b}sin(\omega_b)$ & $\mathcal{U}(-1,1)$ & $0.34^{+0.21}_{-0.30}$ & \text{Parametrization for $e$ and $\omega$} \\
            \noalign{\smallskip}
            $S_{2,b}=\sqrt{e_b}cos(\omega_b)$ & $\mathcal{U}(-1,1)$ & $0.30^{+0.13}_{-0.20}$ & \text{Parametrization for $e$ and $\omega$} \\
            \noalign{\smallskip}
            \hline
            \noalign{\smallskip}
            RV parameters for HARPS \\
            $\sigma_{w,HARPS}$ (km/s) & log$\mathcal{U}(10^{-4},10^{-2})$ & $0.00039^{+0.00078}_{-0.00024}$ & \text{Extra jitter term for HARPS} \\
            \noalign{\smallskip}
            RV slope (km/s/days) & $\mathcal{U}(-1.,1.)$ & $\left(1.81^{+0.86}_{-0.89}\right)\e{-5}$ & \text{Linear trend : first parameter} \\
            \noalign{\smallskip}
            RV intercept (systemic velocity in km/s) & $\mathcal{U}(37.18,37.28)$ & $37.22861^{+0.00092}_{-0.00077}$ & \text{Linear trend : second parameter} \\
            \noalign{\smallskip}
            \hline
         \end{tabular}
\label{tab:1}
\end{table*}

\begin{table*}[t]
      \caption[]{Prior Values for the joint fit of TESS, ExTrA, LCO, and HARPS}\label{table:priorjoint}
         \label{}
         \begin{tabular}{p{0.3\linewidth}cc}
            \hline
            \noalign{\smallskip}
            Parameter Name & \text{Prior} & \text{Description} \\
            \noalign{\smallskip}
            \hline
            \noalign{\smallskip}
            Parameters for TOI-269~b      \\
            $P_b$ (days) & $\mathcal{U}(3.69769,3.69773)$ & \text{Period of the planet}           \\
            $t_0$ (BJD TDB)& $\mathcal{U}(2458381.845,2458381.849)$ & \text{Time of transit-center for the planet} \\
            $\rho_{\star}$ $g/cm^3$ & $\mathcal{N}(8.760,0.870)$ & \text{Stellar density}  \\
            $r_{1,b}$ & $\mathcal{U}(0.3,1)$ & \text{Parametrization for p and b} \\
            $r_{2,b}$ & $\mathcal{U}(0,0.1)$ & \text{Parametrization for p and b} \\
            $S_{1,b}=\sqrt{e_b}sin(\omega_b)$ & $\mathcal{U}(-1,1)$ & \text{Parametrization for $e$ and $\omega$} \\
            $S_{2,b}=\sqrt{e_b}cos(\omega_b)$ & $\mathcal{U}(-1,1)$ & \text{Parametrization for $e$ and $\omega$} \\
            $K_b$ (km/s) & $\mathcal{U}(0,0.016)$ & \text{Radial-velocity semi-amplitude of the planet} \\
            \noalign{\smallskip}
            \hline
            \noalign{\smallskip}
            Parameters for TESS photometry \\
            $D_{TESS}$ & 1.0 & \text{Dilution factor for TESS} \\
            $M_{TESS}$ & $\mathcal{N}(0,0.1^2)$ & \text{Relative flux offset for TESS} \\
            $\sigma_{w,TESS}$ (ppm) & log$\mathcal{U}(1,100)$ & \text{Extra jitter term for TESS lightcurve} \\
            $q_{1,TESS}$ & $\mathcal{U}(0,1)$ & \text{Quadratic limb-darkening parametrization} \\
            $q_{2,TESS}$ & $\mathcal{U}(0,1)$ & \text{Quadratic limb-darkening parametrization}  \\
            \noalign{\smallskip}
            \hline
            \noalign{\smallskip}
            Parameters for LCO-CTIO photometry \\
            $D_{LCO}$ & 1.0 & \text{Dilution factor for LCO-CTIO} \\
            $M_{LCO}$ & $\mathcal{N}(0,0.1^2)$ & \text{Relative flux offset for LCO-CTIO} \\
            $\sigma_{w,LCO}$ (ppm) & log$\mathcal{U}(1,2000)$ & \text{Extra jitter term for LCO-CTIO lightcurve} \\
            $q_{1,LCO}$ & $\mathcal{U}(0,1)$ & \text{Quadratic limb-darkening parametrization} \\
            $q_{2,LCO}$ & $\mathcal{U}(0,1)$ & \text{Quadratic limb-darkening parametrization} \\
            \noalign{\smallskip}
            Parameters of the GP \\
            $\sigma_{GP,LCO}$ (relative flux) & log$\mathcal{U}(10^{-6},1)$ & \text{Amplitude of the GP} \\
            $\rho_{GP,LCO}$ (days) & log$\mathcal{U}(10^{-3},10)$ & \text{Time-scale of the Matern kernel} \\
            \noalign{\smallskip}
            \hline
            \noalign{\smallskip}
            Parameters for ExTrA photometry \\
            $D_{ExTrA}$ & 1.0 & \text{Dilution factor for ExTrA} \\
            $M_{ExTrA}$ & $\mathcal{N}(0,0.1^2)$ & \text{Relative flux offset for each ExTrA light curve} \\
            $\sigma_{w,ExTrA}$ (ppm) & log$\mathcal{U}(1,1000)$ & \text{Extra jitter term for each ExTrA light curve} \\
            $q_{1,ExTrA}$ & $\mathcal{U}(0,1)$ & \text{Quadratic limb-darkening parametrization} \\
            $q_{2,ExTrA}$ & $\mathcal{U}(0,1)$ & \text{Quadratic limb-darkening parametrization} \\
            \noalign{\smallskip}
            Parameters of the GP for each ExTrA light curve (one for each night and each telescope)\\
            $\sigma_{GP,ExTrA}$ (relative flux) & log$\mathcal{U}(10^{-6},1)$ & \text{Amplitude of the GP} \\
            $\rho_{GP,ExTrA}$ (days) & log$\mathcal{U}(10^{-3},10)$ & \text{Time-scale of the Matern kernel} \\
            \noalign{\smallskip}
            \hline
            \noalign{\smallskip}
            RV parameters for HARPS \\
            $\sigma_{w,HARPS}$ (km/s) & log$\mathcal{U}(10^{-4},10^{-2})$ & \text{Extra jitter term for HARPS} \\
            RV slope (km/s/days) & $\mathcal{U}(-10^{-4},10^{-4})$ & \text{Linear trend : first parameter} \\
            RV intercept (systemic velocity) (km/s) & $\mathcal{U}(37.22,37.24)$ & \text{Linear trend : second parameter} \\
            \noalign{\smallskip}
            \hline
         \end{tabular}
\label{tab:1}
\end{table*}

\begin{table*}
      \caption[]{Posterior parameters for the different instruments used in the \juliet analysis for TOI-269~b. The posterior estimate corresponds to the median value. Error bars denote the 68\% posterior credibility intervals.}
      \label{table:postinstru}
         \label{}
         \begin{tabular}{p{0.4\linewidth}cc}
            \hline
            \noalign{\smallskip}
            Parameter name & \text{Posterior estimate} & \text{Description}\\
            \noalign{\smallskip}
            \hline
            \noalign{\smallskip}
            Posterior parameters for TESS photometry \\
            $M_{TESS}$ (ppm)& $0.00033^{+0.00012}_{-0.00012}$ & \text{Relative flux offset}\\
            \noalign{\smallskip}
            $\sigma_{w,TESS}$ (ppm) & $9.6^{+34}_{-7.4}$ & \text{Extra jitter term}\\
            \noalign{\smallskip}
            \hline
            \noalign{\smallskip}
            Posterior parameters for LCO-CTIO photometry \\
            $M_{LCO}$ (ppm) & $0.0009^{+0.0022}_{-0.0020}$ & \text{Relative flux offset}\\
            \noalign{\smallskip}
            $\sigma_{w,LCO}$ (ppm) & $890^{+120}_{-120}$ & \text{Extra jitter term}\\
            \noalign{\smallskip}
            Parameters of the GP \\
            $\sigma_{GP,LCO}$ (relative flux) & $0.0030^{+0.0044}_{-0.0014}$ & \text{Amplitude of the GP}\\
            \noalign{\smallskip}
            $\rho_{GP,LCO}$ (days) & $0.094^{+0.14}_{-0.051}$ & \text{Time-scale of the Matern kernel}\\
            \noalign{\smallskip}
            \hline
            \noalign{\smallskip}
            RV parameters for HARPS \\
            $\sigma_{w,HARPS}$ (m/s) & $0.39^{+0.68}_{-0.24}$ & \text{Extra jitter term}\\
            \noalign{\smallskip}
            \hline
            \noalign{\smallskip}
            Posterior parameters for ExTrA photometry \\
            \noalign{\smallskip}
            Posterior parameters for night 1 of observations \\
            \noalign{\smallskip}
            Telescope 2\\
            $M_{ExTrA}$ (ppm) & $0.00005^{+0.050}_{-0.038}$ & \text{Relative flux offset}\\
            \noalign{\smallskip}
            $\sigma_{w,ExTrA}$ (ppm) & $22^{+160}_{-19}$ & \text{Extra jitter term}\\
            \noalign{\smallskip}
            $\sigma_{GP,ExTrA}$ (relative flux) & $0.059^{+0.095}_{-0.031}$ & \text{Amplitude of the GP}\\
            \noalign{\smallskip}
            $\rho_{GP,ExTrA}$ (days) & $0.23^{+0.23}_{-0.10}$ & \text{Time-scale of the Matern kernel}\\
            \noalign{\smallskip}
            Telescope 3\\
            $M_{ExTrA}$ (ppm) & $-0.005^{+0.013}_{-0.025}$ & \text{Relative flux offset}\\
            \noalign{\smallskip}
            $\sigma_{w,ExTrA}$ (ppm) & $27^{+210}_{-24}$ & \text{Extra jitter term}\\
            \noalign{\smallskip}
            $\sigma_{GP,ExTrA}$ (relative flux) & $0.019^{+0.052}_{-0.013}$ & \text{Amplitude of the GP}\\
            \noalign{\smallskip}
            $\rho_{GP,ExTrA}$ (days) & $0.37^{+0.76}_{-0.25}$ & \text{Time-scale of the Matern kernel}\\
            \noalign{\smallskip}
            \hline
            \noalign{\smallskip}
            Posterior parameters for night 2 of observations \\
            \noalign{\smallskip}
            Telescope 2\\
            $M_{ExTrA}$ (ppm) & $0.0010^{+0.018}_{-0.012}$ & \text{Relative flux offset}\\
            \noalign{\smallskip}
            $\sigma_{w,ExTrA}$ (ppm) & $34^{+290}_{-31}$ & \text{Extra jitter term}\\
            \noalign{\smallskip}
            $\sigma_{GP,ExTrA}$ (relative flux) & $0.0163^{+0.028}_{-0.0079}$ & \text{Amplitude of the GP}\\
            \noalign{\smallskip}
            $\rho_{GP,ExTrA}$ (days) & $0.124^{+0.13}_{-0.052}$ & \text{Time-scale of the Matern kernel}\\
            \noalign{\smallskip}
            Telescope 3\\
            $M_{ExTrA}$ (ppm) & $-0.0008^{+0.0023}_{-0.0017}$ & \text{Relative flux offset}\\
            \noalign{\smallskip}
            $\sigma_{w,ExTrA}$ (ppm) & $29^{+240}_{-26}$ & \text{Extra jitter term}\\
            \noalign{\smallskip}
            $\sigma_{GP,ExTrA}$ (relative flux) & $0.00232^{+0.0075}_{-0.00096}$ & \text{Amplitude of the GP}\\
            \noalign{\smallskip}
            $\rho_{GP,ExTrA}$ (days) & $0.035^{+0.13}_{-0.021}$ & \text{Time-scale of the Matern kernel}\\
            \noalign{\smallskip}
            \hline
            \noalign{\smallskip}
            Posterior parameters for night 3 of observations \\
            \noalign{\smallskip}
            Telescope 1\\
            $M_{ExTrA}$ (ppm) & $-0.00077^{+0.00036}_{-0.00032}$ & \text{Relative flux offset}\\
            \noalign{\smallskip}
            $\sigma_{w,ExTrA}$ (ppm) & $26^{+210}_{-23}$ & \text{Extra jitter term}\\
            \noalign{\smallskip}
            $\sigma_{GP,ExTrA}$ (relative flux) & $0.00067^{+0.00061}_{-0.00061}$ & \text{Amplitude of the GP}\\
            \noalign{\smallskip}
            $\rho_{GP,ExTrA}$ (days) & $0.016^{+0.42}_{-0.012}$ & \text{Time-scale of the Matern kernel}\\
            \noalign{\smallskip}
            Telescope 2\\
            $M_{ExTrA}$ (ppm) & $-0.016^{+0.041}_{-0.039}$ & \text{Relative flux offset}\\
            \noalign{\smallskip}
            $\sigma_{w,ExTrA}$ (ppm) & $30^{+260}_{-27}$ & \text{Extra jitter term}\\
            \noalign{\smallskip}
            $\sigma_{GP,ExTrA}$ (relative flux) & $0.055^{+0.11}_{-0.034}$ & \text{Amplitude of the GP}\\
            \noalign{\smallskip}
            $\rho_{GP,ExTrA}$ (days) & $0.33^{+0.41}_{-0.17}$ & \text{Time-scale of the Matern kernel}\\
            \noalign{\smallskip}
            Telescope 3\\
            $M_{ExTrA}$ (ppm) & $0.0053^{+0.018}_{-0.0098}$ & \text{Relative flux offset} \\
            \noalign{\smallskip}
            $\sigma_{w,ExTrA}$ (ppm) & $27^{+220}_{-24}$ & \text{Extra jitter term}\\
            \noalign{\smallskip}
            $\sigma_{GP,ExTrA}$ (relative flux) & $0.0160^{+0.026}_{-0.0073}$ & \text{Amplitude of the GP}\\
            \noalign{\smallskip}
            $\rho_{GP,ExTrA}$ (days) & $0.071^{+0.076}_{-0.029}$ & \text{Time-scale of the Matern kernel}\\
            \noalign{\smallskip}
            \hline
         \end{tabular}
\label{tab:1}
\end{table*}

\clearpage

\section{Spectral energy distribution}\label{section:sed}

\begin{figure}
  \centering
  \includegraphics[width=0.5\textwidth]{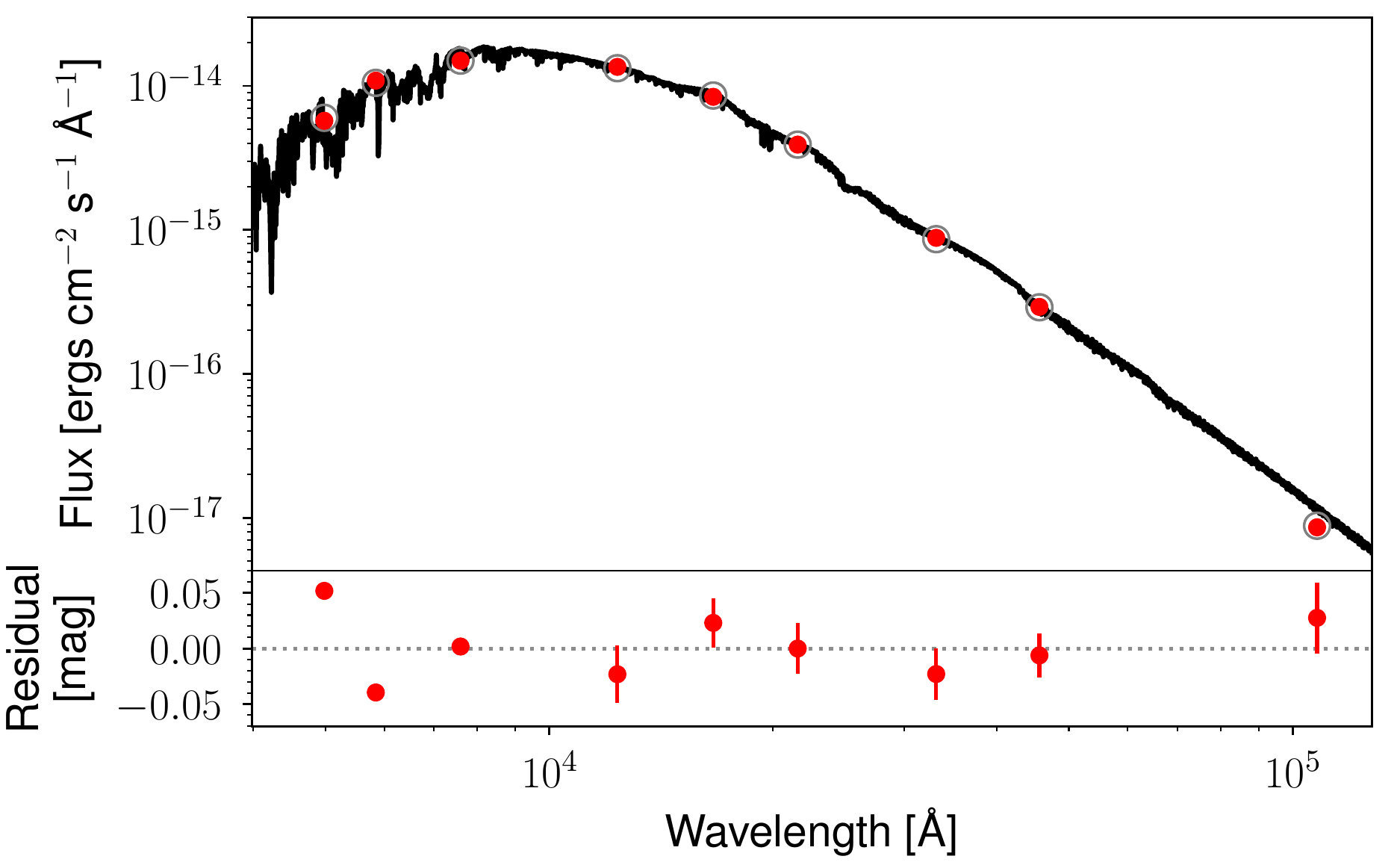}
  \caption{Spectral energy distribution of TOI-269. The solid line is the MAP PHOENIX/BT-Settl interpolated synthetic spectrum, red circles are the absolute photometric observations, and gray open circles are the result of integrating the synthetic spectrum in the observed band-passes.} \label{fig:sed}
\end{figure}

We performed an independent determination of the stellar parameters of TOI-269 modeling the spectral energy distribution (SED) with stellar atmosphere and evolution models. We constructed the SED using the magnitudes from Gaia DR2 \citep{evans2018, maiz2018}, the 2-Micron All-Sky Survey \citep[2MASS,][]{2mass,cutri2003}, and the Wide-field Infrared Survey Explorer \citep[WISE,][]{wise,cutri2013}. We modeled these magnitudes using the procedure described in \citet{diaz2014}. We used informative priors for the effective temperature ($T_{\mathrm{eff}}$), and metallicity ($[\rm{Fe/H}]$) from the analysis of the HARPS co-added spectra with {\sc \tt SpecMatch-Emp} \citep{yee2017}, and for the distance from Gaia DR2 \citep{gaia2018,bailer-jones2018}. We used non-informative priors for the rest of the parameters. We used the PHOENIX/BT-Settl stellar atmosphere models \citep{allard2012}, and two stellar evolution models: Dartmouth \citep{dotter2008} and {\sc \tt PARSEC} \citep{chen2014}. We did three analyses, one using stellar atmosphere models only (BT-Settl) and another two with the stellar atmosphere models and each of the stellar evolution models (BT-Settl + Dartmouth, BT-Settl + {\sc \tt PARSEC}). The priors, posterior median, and 68.3\% credible intervals (CI) for jump and derived parameters are listed in Table~\ref{table:sed}. The data with the maximum a posteriori (MAP) stellar atmosphere model is shown in Fig.~\ref{fig:sed}.

\begin{table*}
    \scriptsize
    \renewcommand{\arraystretch}{1.25}
    \setlength{\tabcolsep}{2pt}
\centering
\caption{Modeling of the spectral energy distribution: Parameter, prior, posterior median, and 68.3\% CI for each of the analyses.}\label{table:sed}~
\begin{tabular}{lccccc}
\hline
Parameter & & Prior & Posterior   & Posterior & Posterior\\
&  & &  BT-Settl & BT-Settl + Dartmouth & BT-Settl + {\sc \tt PARSEC}\\
\hline
Effective temperature, $T_{\mathrm{eff}}$ & [K]     & $N$(3514, 70)             & 3578$^{+44}_{-38}$        & 3586$^{+24}_{-20}$              & 3562$\pm$37                     \\
Surface gravity, $\log g$                   & [cgs]   & $U$(-0.5, 6.0)            & 5.46$^{+0.37}_{-0.52}$    & 4.8648$\pm$0.0055               & 4.833$^{+0.021}_{-0.036}$       \\
Metallicity, $[\rm{Fe/H}]$                & [dex]   & $N$(-0.29, 0.12)          & -0.32$\pm$0.11            & -0.325$\pm$0.082                & -0.403$^{+0.073}_{-0.085}$      \\
Distance                                  & [pc]    & $N$(57.023, 0.076)        & 57.023$\pm$0.076          & 57.022$\pm$0.077                & 57.022$\pm$0.079                \\
$E_{\mathrm{(B-V)}}$                      & [mag]   & $U$(0,3)                  & 0.028$^{+0.036}_{-0.020}$ & 0.025$^{+0.030}_{-0.018}$       & 0.022$^{+0.030}_{-0.016}$       \\
Jitter Gaia                               & [mag]   & $U$(0, 1)                 & 0.069$^{+0.10}_{-0.034}$  & 0.075$^{+0.098}_{-0.034}$       & 0.082$^{+0.12}_{-0.040}$        \\
Jitter 2MASS                              & [mag]   & $U$(0, 1)                 & 0.025$^{+0.043}_{-0.018}$ & 0.030$^{+0.045}_{-0.021}$       & 0.026$^{+0.044}_{-0.018}$       \\
Jitter WISE                               & [mag]   & $U$(0, 1)                 & 0.025$^{+0.044}_{-0.018}$ & 0.023$^{+0.040}_{-0.017}$       & 0.029$^{+0.053}_{-0.021}$       \\
Radius, $R_\star$                         & [R$_\odot$] &  $U$(0, 100)$\dagger$ & 0.3926$\pm$0.062          & 0.3910$\pm$0.0032               & 0.3948$^{+0.0064}_{-0.0052}$    \\
Mass, $M_\star$                           & [M$_\odot$] &                       &                           & 0.4103$\pm$0.0045               & 0.390$^{+0.014}_{-0.023}$       \\
Density, $\rho_\star$                     & [$\mathrm{g\;cm^{-3}}$] &           &                           & 9.670$\pm$0.180                    & 8.920$^{+0.540}_{-0.800}$            \\
Age                                       & [Gyr]       &                       &                           & 5.3$^{+3.5}_{-2.4}$             & 0.143$^{+0.15}_{-0.049}$        \\
Luminosity                                & [L$_\odot$] &                       &                           & 0.02276$^{+0.00063}_{-0.00050}$ & 0.02261$^{+0.00066}_{-0.00054}$ \\
\hline
\end{tabular}
\begin{list}{}{}
\item {\bf{Notes.}}
$N$($\mu$,$\sigma$): Normal distribution prior with mean $\mu$, and standard deviation $\sigma$. $U$(l,u): Uniform distribution prior in the range [l, u]. $\dagger$ For the BT-Settl analysis only.
\end{list}
\end{table*}

\end{appendix}

\end{document}